\documentclass[showpacs,pre,superscriptaddress,10pt]{revtex4-2}

\usepackage{natbib}
\usepackage{braket}
\usepackage{amsmath}
\usepackage{amsfonts}
\usepackage{amssymb}
\usepackage{color}
\usepackage{xcolor}
\usepackage{mathtools}
\usepackage{mathrsfs}
\usepackage[normalem]{ulem}
\usepackage{appendix}
\usepackage{graphicx}

\usepackage{hyperref}
\begin{document}
\title{Changing the order of a dynamical phase transition through fluctuations in a quantum p-spin model}
\author{Lorenzo Correale}
\affiliation{SISSA --- International School for Advanced Studies, via Bonomea 265, I-34136 Trieste, Italy}
\affiliation{INFN --- Istituto Nazionale di Fisica Nucleare, Sezione di Trieste, I-34136 Trieste, Italy}
\author{Alessandro Silva}
\affiliation{SISSA --- International School for Advanced Studies, via Bonomea 265, I-34136 Trieste, Italy}

\begin{abstract}
We study the non-equilibrium phase diagram of a fully-connected Ising $p$-spin model, for generic $p>2$, and investigate its robustness with respect to the inclusion of spin-wave fluctuations, resulting from a ferromagnetic, short-range spin interaction.
In particular, we investigate the dynamics of the mean-field model after a quantum quench: we observe a new dynamical phase transition which is either first or second order depending on the even or odd parity of $p$, in stark contrast with its thermal counterpart which is first order for all $p$. 
The dynamical phase diagram is qualitatively modified by the fluctuations introduced by a short-range interaction which drive the system always towards
various prethermal paramagnetic phases determined by the strength of time dependent fluctuations of the magnetization.
\end{abstract}



\date{\today}
   \maketitle

\section{Introduction}
Equilibrium phase transitions, either at zero or finite-temperature, are known to leave a substantial imprint in the non-equilibrium dynamics of a quantum many-body system~\cite{Polkovnikov2011}. For example, even when a stationary state attained after a quantum quench does not reveal  signatures of order as in low-dimensional systems ~\cite{Rossini09,Calabrese11,Maraga14}, a linear ramp through a second order quantum critical point leaves universal signatures in the scaling of the number of excitations with the ramp speed~\cite{Polkovnikov05,Zurek05,Dziarmaga05,Cherng06,King2022JZ,Du2023KZ,Bando2023KZ}, as confirmed extensively in a number of experiments~\cite{Braun15,Anquez16,Clark16,Clark17,Kang17,Keesling19,Cui20}. Analogous signatures are left when a first-order quantum phase transition is crossed~\cite{Swislocki13,Coulamy17,Shimizu18} through the nucleation of resonant bubbles of the new phase close to the critical point~\cite{DelRe16,Sinha21,lagnese21} which leads to a modified Kibble-Zurek-like power-law scaling~\cite{Qiu21}.   

Among the signatures of criticality observed out-of-equilibrium, Dynamical  Phase Transitions (DPT) occupy a special place. A dynamical quantum criticality can be observed as singular temporal behavior of the Loschmidt echo (LE) most notably after a quench across a quantum phase transition~\cite{heyl2013dynamical,Heyl2018,Weidinger17}, even in situations where long-range order cannot be sustained in stationary states. In systems with long-range interactions, on the contrary, intertwined with the singular behavior of the LE~\cite{Zunkovic18,Halimeh2017,lang2018dynamical,lang2018concurrence,halimeh2016dynamical,homrighausen2017anomalous}, a standard Landau-type critical behavior based on the dependence of a time-averaged order parameter with respect to the quench parameters can be observed~\cite{Sciolla2011,Jurcevic17}.
Peculiar to the second-order dynamical transitions arising in these models is the fact that they are associated to critical trajectories
with a divergent time scale in the dynamics separating
revivals with a finite order parameter ~\cite{Zunkovic18}.
In the presence of fluctuations critical trajectories become unstable and second order dynamical critical points widen up into chaotic dynamical phases~\cite{Lerose2018,Lerose2019,Piccitto19,Piccitto19b}. \\

While a great deal is known about second-order dynamical phase transitions, the dynamics of systems displaying equilibrium first order
transitions is much less explored.
The notion of dynamical criticality associated to the LE has been extended to include first order behavior~\cite{Canovi14} while first-order and dissipative phase 
transitions in infinite range $p$-spin systems coupled to an external bath have been studied in Ref.[\onlinecite{wang2021dissipative}].
However, dynamical transitions occurring in systems displaying first order equilibrium transitions are much less studied.

In this work we address this issue by studying
dynamical phase transitions and their stability against fluctuations in a system displaying a first order equilibrium transition: a spin system with infinite range $p$-spin interactions in a transverse field. We show that, already at mean-field level where the dynamics is  effectively classical, the system undergoes a DPT after a quench of the transverse field $g$, whose order depends non-trivially on $p$, despite its equilibrium counterpart being always of first order. In particular, we show that the order of the dynamical transition can be inferred entirely from the profile of the  underlying energy landscape.
We then perturb the model by a short-range two-body interaction tuning the strength of spin fluctuations \cite{Lerose2018,Lerose2019,ruckriegel2012time}. While for $p=2$ a chaotic dynamical region opens up near mean-field criticality \cite{Lerose2018,Lerose2019}, we show that for $p>2$ dynamical chaos is almost entirely replaced by a new prethermal regime, which we define as ``dynamical paramagnetic phase", which appears for sufficiently large short-range coupling. This is due to the emission of energy in the form of spin-waves, which predominantly drive the system into a paramagnetic minimum even in the presence of other minima in the energy landscape.\\

\section{Mean-field dynamics}\label{SEC_mean-field_dynamics}

In this section, we study the dynamics of $N$ spins $ 1/2$ subject to all-to-all $p$-body and a global transverse field $\Tilde{g}$. The corresponding Hamiltonian is given by:
\begin{equation} \label{eq:pspin_mf_hamiltonian}
    \hat{H}_0 = -\frac{\lambda}{2N^{p-1}} \sum_{i_1\dots i_p =1}^N \hat{\sigma}^x_{i_1} \dotsc \hat{\sigma}^x_{i_p} - \frac{\tilde{g} }{2} \sum_i \hat{\sigma}^z_i \ .
\end{equation}
Here, the operators $\hat{\sigma}^\alpha_i$ denote the Pauli matrices at site $i$. The fully connected $p$-spin model in Eq.~\eqref{eq:pspin_mf_hamiltonian}, which was originally introduced in the context of spin glasses~\cite{derrida1980random,derrida1981random}, plays a central role for studies on quantum annealing \cite{jorg2010energy,bapst2012quantum}. Its zero-temperature equilibrium phase diagram can be derived analytically \cite{jorg2010energy,bapst2012quantum, das2006infinite, filippone2011quantum}, utilizing  a Suzuki-Trotter decomposition in the thermodynamic limit~\cite{sachdev2011quantum}.
The phase diagram displays a quantum phase transition driven by $\Tilde{g}$ and detected by the magnetization $\mathcal{S}^x=\sum_i\braket{\hat{\sigma}_i^x}/N$ along the $x$-axis. The transition, located at some $\Tilde{g}=g_c$, separates a ferromagnetic state, defined by a non-vanishing $\mathcal{S}^x$, from a paramagnetic where $\mathcal{S}^x=0$. The transition is continuous for $p=2$, where $\mathcal{S}^x$ vanishes with a square-root singularity~\cite{das2006infinite}, while it is of the first order for $p>2$, where $\mathcal{S}^x$ displays a discontinuity. In this section we will address the dynamics and dynamical phase transitions of this $p$-spin model.\\

\begin{figure*}[ht!]
\includegraphics[width = \textwidth]{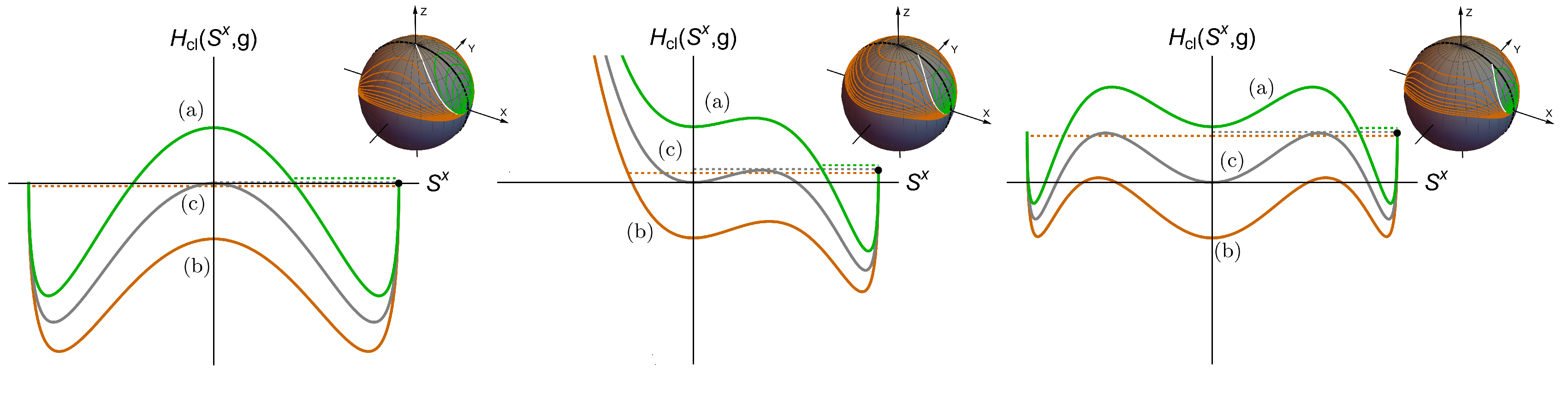}
\centering
\caption{(Color online) Energy profiles described by Eq.~\eqref{energylandscape}, projected on the plane $\mathcal{S}^y=0$. The projection is obtained from the black cut on the Northern hemisphere of the corresponding Bloch spheres, where $\mathcal{S}^z = \sqrt{1-(\mathcal{S}^x)^2}$.
Each panel illustrates the shape of the energy landscape corresponding to a different set of values of $p$: $p=2$ (left), $p=2n+1$ (center) and $p=2n+2$ (right), respectively, with $n \geq 1$ integer. For practical purposes, we fixed $n=1$. Each profile in each panel corresponds to a different value of $g$: $g=g_{dyn}-0.1$ (a), $g=g_{dyn}+0.1$ (b) and $g=g_{dyn}$ (c). The value of $g_{dyn}$ is obtained from Eqs.~\eqref{eq:pspin_clean_red_stat_eqs} and~\eqref{eq:pspin_clean_equate_energies}, as discussed in the main text. For each energy profile, the corresponding trajectory evolving from $\vec{\mathcal{S}}(0)=\mathbf{x}$ (black dot on the profile) is represented by a dashed horizontal line. In particular, the gray dashed horizontal line corresponds to a separatrix (white trajectory on the Bloch sphere). See also the orbits in Fig. \ref{fig:meanfieldresults} (a) and (d) for a comparison.}
\label{fig:pspin_clean_landscape}
\end{figure*} 

\subsection{Semi-classical theory for the post-quench dynamics}\label{sec_sciolla_theory}

Our goal is to study the dynamics of the average magnetization, $\vec{\mathcal{S}}(t) = \braket{\sum_j \vec{\sigma}_j(t)}/N$, after a quench in the transverse field $\tilde{g}$ in Eq.~\eqref{eq:pspin_mf_hamiltonian}, where the field $g$ abruptly changes from a value $g_0=0$ to a value $g>0$. Specifically, we prepare the system in the fully polarized state,
\begin{equation}\label{eq:initial_state}
\ket{\psi_0}=\ket{\rightarrow\cdots\rightarrow} ~ ,
\end{equation}
corresponding to a ground state $\hat{H}_0$ at $g_0=0$.
The dynamics of $\vec{\mathcal{S}}(t)$ is obtained by averaging over $\ket{\psi_0}$ the Heisenberg equations generated by the Hamiltonian $\hat{H}_0$.
However, due to the all-to-all interacting nature of $\hat{H}_0$, this dynamics becomes effectively classical in the thermodynamic limit $N\to\infty$. This classical behavior is derived from the general theory outlined in Ref.~\cite{Sciolla2011}, which we briefly review in the following.\\

To begin with, we observe that the Hamiltonian in Eq.~\eqref{eq:pspin_mf_hamiltonian}, for the post-quench transverse field $\Tilde{g}=g$, can be rewritten in terms of the collective spin operators, $\hat{S}^\alpha=\sum_j \hat{\sigma}_j^\alpha/N$, as follows:
\begin{equation}
    \hat{H}_0 = -N \left[ \frac{\lambda}{2}(\hat{S}^x)^p + \frac g2 \hat{S}^z \right] ~ .
\end{equation}
The collective spin components satisfy the commutation relations
\begin{equation}\label{eq:coll_spin_commutators}    \left[\hat{S}^\alpha,\hat{S}^\beta\right] = \frac{2i}{N}\sum_{\gamma=x,y,z} \epsilon^{\alpha\beta\gamma}\hat{S}^\gamma   \ ,  
\end{equation}
where $\epsilon^{\alpha\beta\gamma}$ is the Levi-Civita symbol. The commutators, controlled by an effective Planck constant $1/N$, vanish in the thermodynamic limit. As a consequence, the dynamics of the average magnetization becomes effectively classical for $N\to\infty$ and is governed by the Hamilton equation (see also~\cite{munoz2020simulation,wang2021dissipative}):
\begin{equation}\label{eq:Hamilton_equation}
    \frac{d \vec{\mathcal{S}}(t)}{dt} = \{\vec{\mathcal{S}}(t), \mathcal{H}_{cl}(\vec{\mathcal{S}}(t),g) \} = 
    \vec{\mathcal{S}}(t) \times   \frac{\partial\mathcal{H}_{cl}(\vec{\mathcal{S}}(t),g) }{\partial \vec{\mathcal{S}} } ~ .
\end{equation}
The right-hand side of Eq.~\eqref{eq:Hamilton_equation} is derived by substituting the rescaled average commutators, $N\braket{[\hat{H}_0,\hat{S}^\alpha]}/i$, with the corresponding Poisson brackets,
$\{\mathcal{H}_{cl}(\vec{\mathcal{S}}), \mathcal{S}^\alpha \}$. Here, the effective \emph{classical} potential $\mathcal{H}_{cl}(\vec{\mathcal{S}},g)$ is obtained as the thermodynamic limit of $\braket{\hat{H}_0}/N$ and is given by
\begin{equation} \label{energylandscape}
\mathcal{H}_{cl}(\vec{\mathcal{S}},g) = - \lambda (\mathcal{S}^x)^p-g \mathcal{S}^z ~ .
\end{equation}
The initial condition for Eq.~\eqref{eq:Hamilton_equation} is $\vec{\mathcal{S}}(0) = \mathbf{x}$, corresponding to the fully polarized state in Eq.~\eqref{eq:initial_state}. Here and throughout this paper, we use the notation $\{\mathbf{x},\mathbf{y},\mathbf{z}\}$ to denote the unit vectors along the corresponding axes.
We also observe that, as the modulus of the magnetization is a constant of motion, the classical dynamics from Eq.~\eqref{eq:Hamilton_equation} takes place on the Bloch sphere $|\vec{\mathcal{S}}|^2=1$. Thus, the system under consideration is always in a non-equilibrium state, as all the microscopic spins perform a coherent, undamped precession. \\

The dynamics from Eq.~\eqref{eq:Hamilton_equation} is strongly influenced by the shape of the effective Hamiltonian $\mathcal{H}_{cl}(\vec{\mathcal{S}},g)$. 
Depending on the value of $p$ and for sufficiently small values of $g$, the profile $\mathcal{H}_{cl}(\vec{\mathcal{S}},g)$ on the Bloch sphere exhibits various topologies, characterized by the number and positions of its maxima and minima. 
To determine their locations, we parameterize the magnetization with the spherical angles  $(\theta,\phi) \in [0,\pi/2] \times [0,2 \pi]$, as $\vec{\mathcal{S}}= \big( \sin \theta \cos \phi, \sin \theta \sin \phi, \cos \theta \big)$. The stationary points of $\mathcal{H}_{cl}(\vec{\mathcal{S}},g)$ are then defined by the equations:
\begin{equation}\label{eq:pspin_clean_stat_points}
\left\{
\begin{split}
    \frac{\partial \mathcal{H}_{cl}}{\partial \phi} &= -\lambda p (\sin \theta)^p (\cos\phi)^{p-1} \sin \phi = 0 \\
    \frac{\partial \mathcal{H}_{cl}}{\partial \theta} &= -\lambda p (\sin\theta)^{p-1} (\cos\phi)^p \cos\theta + g \sin \theta = 0 \ . \\ 
\end{split}
\right.
\end{equation}
In the following, consider only the stationary points falling in the in the Northern hemisphere ($\mathcal{S}^z>0$) of the Bloch sphere, where the dynamics is confined for $g>0$.
One possible solution of the system of Eqs.~\eqref{eq:pspin_clean_stat_points} is given by $\theta=0$ (the North Pole $\vec{\mathcal{S}} = \mathbf{z}$ of the Bloch sphere), being a maximum for $p=2$ and a minimum for $p>2$. All the other solutions are obtained by solving the system:
\begin{equation} \label{eq:pspin_clean_red_stat_eqs}
    \left\{
    \begin{split}
        &\sin \phi = 0 \\
        &(\sin\theta)^{p-2} \cos \theta (\cos\phi)^p= \frac g{\lambda p} ~ .
    \end{split}
    \right.
\end{equation}
From the first equation, we get that stationary points lie in the plane $\mathcal{S}^y = 0$ for every $p$, while the number and the precise location of the solutions depend on the value of $p$.
Specifically, for $p=2$, we observe two symmetric minima separated by a maximum at the North pole, $\vec{\mathcal{S}} = \mathbf{z}$. For $p\geq 3$ odd, the same topology persists, but the profile becomes asymmetric with respect to the North pole. Conversely, for $p\geq 4$ even, the potential features three minima: one at the North pole and two symmetric minima with respect to it. To summarize, the projection of $\mathcal{H}_{cl}(\vec{\mathcal{S}},g)$ in the plane $\mathcal{S}^y=0$ exhibits three distinct shapes, corresponding to different values of $p$:
\begin{itemize}
\item[-] A symmetric double-well for $p=2$.
\item[-] An asymmetric double-well for $p \geq 3$ (odd), with one paramagnetic and one ferromagnetic minimum.
\item[-] A symmetric triple-well for $p \geq 4$ (even), with one paramagnetic minimum and two opposite ferromagnetic minima.
\end{itemize}

It is worth noting that these three profiles undergo a qualitative transformation beyond the \emph{spinodal point}~\cite{bapst2012quantum}, defined as
\begin{equation}\label{eq:spinodal_point}
g_{sp}=p(p-2)^{(p-2)/2}/(p-1)^{(p-1)/2} \ .
\end{equation}
More precisely, for each value of $p$ the profile of $\mathcal{H}_{cl}(\vec{\mathcal{S}},g)$ becomes a single well, centered around its only minimum at $\vec{\mathcal{S}} = \mathbf{z}$, when $g>g_{sp}$. For these values of $g$, the second of Eqs.~\eqref{eq:pspin_clean_red_stat_eqs} has no solution.

\subsection{The dynamical transition}

The presence of multiple local minima in the profile has a strong impact on the dynamics of $\vec{\mathcal{S}}(t)$, which evolves according to Eq.~\eqref{eq:Hamilton_equation} from the initial condition $\vec{\mathcal{S}}(0)=\mathbf{x}$, located in the rightmost ferromagnetic well. We denote the position of the local maximum that separates this well from the rest of the landscape as $\vec{\mathcal{S}}_m(g)$.
The dynamics exhibits qualitatively different orbits, depending on the value of the post-quench transverse field $g$~\footnote{Our discussion can also be generalized to non-vanishing pre-quench values of the transverse field, $g_0>0$, following the approach of Ref. \cite{Sciolla2011}.}, as qualitatively depicted in Fig.~\ref{fig:pspin_clean_landscape}:
\begin{enumerate}
    \item Below a certain threshold, $g<g_{dyn}$ (trajectory (a)), the dynamics starts from a ferromagnetic well, and the initial energy is insufficient to surmount the energy barrier in correspondence of nearest maximum in $\vec{\mathcal{S}}= \vec{\mathcal{S}}_m(g)$. Consequently, the magnetization oscillates in the ferromagnetic well, with $\mathcal{S}^x(t)>\mathcal{S}^x_m(g)$ at every time $t$.

    \item For $g>g_{dyn}$ (trajectory (b)), the post-quench energy is larger then the energy barrier and the corresponding orbit encompasses all the minima in the landscape.

    \item At precisely $g=g_{dyn}$ (trajectory (c)), the system has sufficient energy to reach the top of the barrier in $\vec{\mathcal{S}}_m(g)$, but it is unable to surpass it. Here the period of the oscillations, $T_{cl}(g)$, diverges and the magnetization approaches $\vec{\mathcal{S}}_m(g)$ infinitely slowly. The resulting orbit forms a \emph{separatrix}.
\end{enumerate}
The three cases listed above correspond to three different possible topologies for the underlying orbits at large times. We classify each distinct topology as a \emph{dynamical phase}~\cite{keeling2010collective}. Consequently, the singular dynamics at $g = g_{dyn}$ leads to a \emph{dynamical phase transition} (DPT).
The corresponding dynamical critical point $g_{dyn}$ is obtained by equating the energy of the initial configuration, $\vec{\mathcal{S}}(0)=\mathbf{x}$, with that of the local maximum in $\vec{\mathcal{S}}_c = \vec{\mathcal{S}}_m(g_{dyn})$. In terms of the variables $\theta$ and $\phi$, the equation reads:
\begin{equation}\label{eq:pspin_clean_equate_energies}
    -\lambda(\sin\theta)^p - g \cos\theta = -g ~ .
\end{equation}
Then, the simultaneous solutions of Eqs.~\eqref{eq:pspin_clean_red_stat_eqs} and~\eqref{eq:pspin_clean_equate_energies} determines both $g_{dyn}$ and the spherical coordinates of the maximum in $\vec{\mathcal{S}}_c$.
From a practical point of view, the DPT can be studied also in terms of a "dynamical order parameter", such as the time-averaged longitudinal magnetization 
\begin{equation}\label{eq:time_avg_magn}
    \overline{\mathcal{S}^x} = \frac 1T \lim_{T\to\infty} \int_0^T dt \mathcal{S}^x(t)  \ .
\end{equation}
As discussed also in Appendix~\ref{APP_classical_period}, the two indicators are equivalent to each other: as $g$ approaches $g_{dyn}$, the change of topology in the orbits is signalled by a divergence of their period $T_{cl}(g)$, in turn creating a non-analiticity in the function $\overline{\mathcal{S}^x}(g)$.
In particular, both $T_{cl}(g)$ and $\overline{\mathcal{S}^x}(g)$ display a log-singularity while approaching the dynamical transition, a feature already known for other mean-field models driven away from equilibrium~\cite{Sciolla2011,Gambassi2011}. It is important to note that such a DPT can be observed in the $N\to\infty$ limit only: for a finite size, quantum fluctuations are restored and induce dephasing between local spins,
leading to the relaxation of time-averaged observables to their thermodynamic expectation values~\cite{Lerose2019}.
In the following, we show that the nature of the mean-field DPT is determined by the topology of the landscape, which depends solely on the parity of $p$ (if $p>2$).
As a consequence, we will focus on studying Eq.~\eqref{eq:Hamilton_equation} for $p=3$ and $p=4$ only. These cases, alongside with 
$p=2$, are paradigmatic and encapsulate the three possible landscape shapes discussed in the previous section, respectively.
\\

\begin{figure*} [t]
\centering
\includegraphics[width = \textwidth]{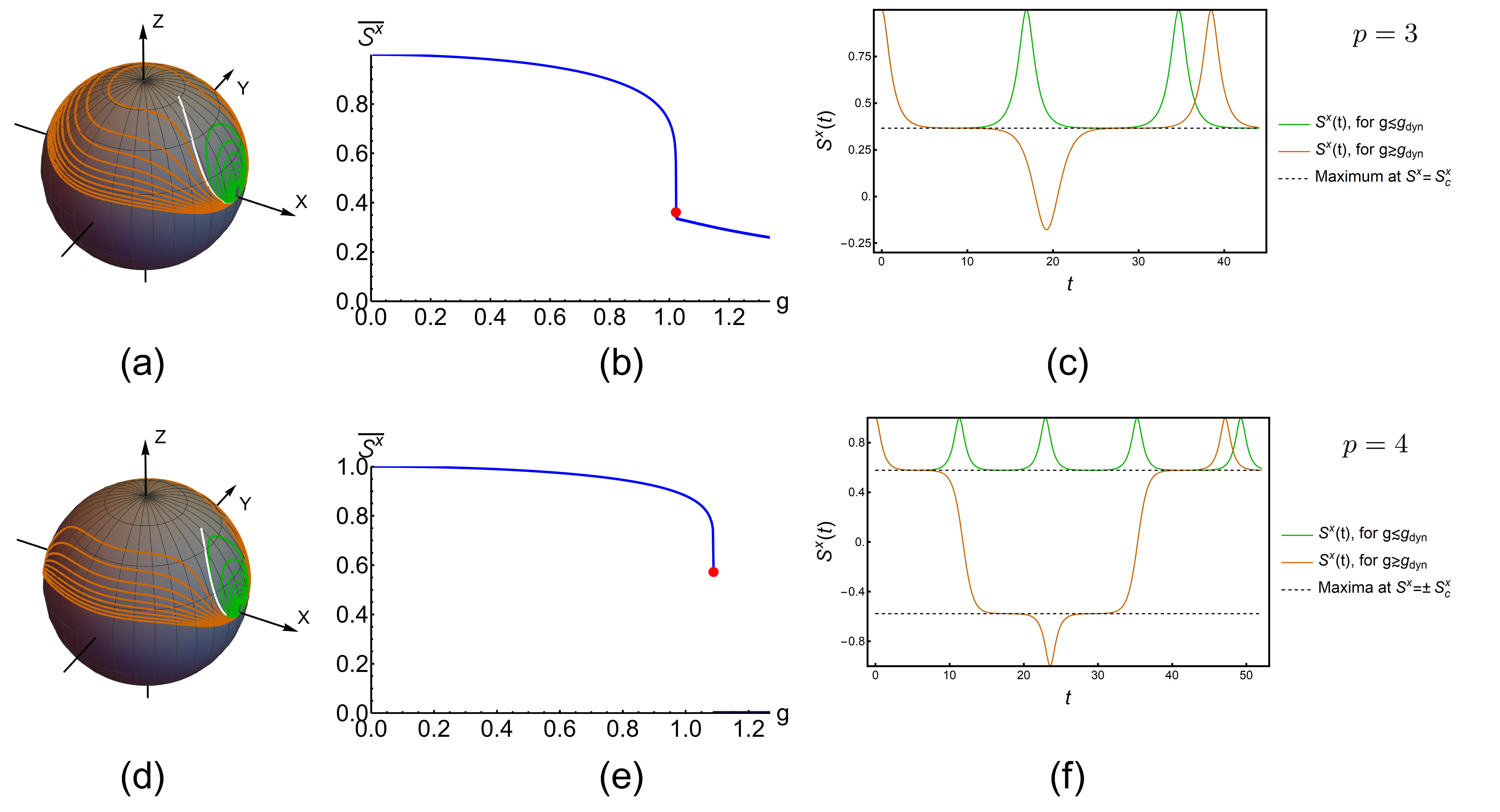}
\caption{ (Color online) 
From left to right: orbits of the magnetization on the Bloch sphere ((a) and (d)),  time-averaged magnetization $\overline{\mathcal{S}^x}$ calculated from Eq.~\eqref{eq:time_avg_magn} as a function of $g$ ((b) and (e)), time evolution  of $\mathcal{S}^x(t)$ near the critical point ((c) and (f)). The plots in the upper row corresponds to the case of $p=3$, the ones below to $p=4$.
\textbf{(Left)} Orbits on the Bloch sphere, resulting from the integration Eq.~\eqref{eq:Hamilton_equation}, from the initial condition $\Vec{\mathcal{S}}=\mathbf{x}$. Each trajectory corresponds to one of $20$ values of $g$,uniformly sampled in the interval $[0.2,3.6]$, where $g_{dyn}$ falls for both $p=3$ and $p=4$.
For both values of $p$, the system evolves along trajectories that are either confined in a ferromagnetic well (green trajectories), for $g<g_{dyn}$, or encompass all the local minima (orange trajectory), for $g>g_{dyn}$. The two regions are divided by a separatrix (white), corresponding to $g=g_{dyn}$. $g_{dyn}$ is calculated from Eqs.~\eqref{eq:pspin_clean_red_stat_eqs} and~\eqref{eq:pspin_clean_equate_energies},as discussed in the main text. 
\textbf{(Center)} Plot of the dynamical order parameter $\overline{\mathcal{S}^x}$, from Eq.~\eqref{eq:time_avg_magn}, as a function of $g$. The red dot correspond to the dynamical critical point $(g_{dyn}, \mathcal{S}^x_c)$, obtained by simultaneously solving simultaneously solving Eqs.~\eqref{eq:pspin_clean_red_stat_eqs} and~\eqref{eq:pspin_clean_equate_energies}.At the dynamical critical point, $\overline{\mathcal{S}^x}$ is continuous for $p=3$ only. 
\textbf{(Right)} Time evolution of the longitudinal magnetization $\mathcal{S}^x(t)$, for a transverse field $g$ either equal to $g_{dyn}-10^{-7}$ (green plots) or to $g_{dyn}+10^{-7}$ (orange plots). For each value of $p$, the dashed lines correspond to the local maxima of the landscape when $g=g_{dyn}$. For each value of $p$, $g_{dyn}$ and $\mathcal{S}^x_c$ are obtained as in the central panels.}
\label{fig:meanfieldresults}
\end{figure*}

First, it is crucial to note that the order of the DPT is not necessarily identical to that of the thermal phase transition for the $p$-spin model at the same $p$ value. The case of $p=2$ is special, exhibiting both a second-order thermal phase transition~\cite{dutta2001phase} and a second-order dynamical phase transition~\cite{Sciolla2011,Zunkovic18,das2006infinite}.
For $p=2$, the dynamical transition occurs at $g=\lambda$ and is identified through $\overline{\mathcal{S}^x}$, which is positive when $g<\lambda$ (indicative of a dynamics confined within a ferromagnetic well) and vanishing for $g>\lambda$ (due to oscillations between the two symmetric wells). We classify these two behaviours as \emph{dynamical
ferromagnetic} and \emph{dynamical paramagnetic} phases, respectively.
The continuity of this DPT is not just a consequence of the symmetry of $\mathcal{H}_{cl}(\vec{\mathcal{S}},g)$ under the reflection $\mathcal{S}^x \to - \mathcal{S}^x$. Instead, it emerges as a general property linked to the topology of the phase space, which is characterized by only one maximum.
To understand why, we notice that,
as $g$ approaches $g_{dyn}$ either from above or below, the energy of the orbit gets close to the one of the separatrix. The separatrix has a crossing at the only local maximum of $\mathcal{H}_{cl}(\vec{\mathcal{S}},g_{dyn})$. Thus, every dynamics starting on the separatrix approaches asymptotically this maximum in $\vec{\mathcal{S}}_c$ ($\vec{\mathcal{S}}_c = \mathbf{z}$ for $p=2$), i.e. 
\begin{equation}
       \lim_{t \to \infty} \vec{\mathcal{S}}(t) = \vec{\mathcal{S}}_c ~ .
\end{equation}
The orbits asymptotically close to the separatrix, approached as $g \to g_{dyn}^\pm$, exhibit a plateau at the only local maximum in $\vec{\mathcal{S}}_c$, whose length diverges with the period $T_{cl}(g)$.  As a consequence, we have that
\begin{equation}
    \lim_{g \to g_{dyn}^\pm} \overline{\mathcal{S}^x}(g) = \mathcal{S}^x_c ~ 
\end{equation}
and the dynamical transition is continuous.\\

For $p>2$, the results illustrated Fig.~\ref{fig:meanfieldresults} consistently demonstrate the occurrence of a DPT at some $g=g_{dyn}$, whose order depends on the parity of $p$, unlike its static counterpart.
For $p=3$ the topology of the phase space is the one of a double well, akin to the case of $p=2$ but without $\mathbb{Z}_2$ symmetry. Consequently, the transition is continuous, following the same argument made for $p=2$ and as illustrated in Fig.~\ref{fig:meanfieldresults}-(b).
Specifically, a dynamical ferromagnetic phase re-emerges for $g<g_{dyn}$, as all the orbits are confined in a ferromagnetic well (green trajectories, Fig.\ref{fig:meanfieldresults}-(a)). Conversely, for $g>g_{dyn}$, the oscillations of the orbits between the two asymmetric wells (orange trajectories, Fig.\ref{fig:meanfieldresults}-(a)) lead to a \emph{dynamical bistable phase}. In this phase, $\overline{\mathcal{S}^x}$ is a weighted average of the two minima, that varies continuously with $g$ and exhibits a cusp at $g = g_{dyn}$.
The connection between the continuity of $\overline{\mathcal{S}^x}$ to the presence of a single local maximum is quantitatively validated by the results presented in Fig.~\ref{fig:meanfieldresults}-(c). This figure illustrates the plateaus developed by the longitudinal magnetization near $g_{dyn}$, both for $g\lesssim g_{dyn}$ (green plot) and $g\gtrsim g_{dyn}$ (orange plot).\\
 
For $p=4$ the DPT becomes discontinuous, as shown in Fig.~\ref{fig:meanfieldresults}-(e). The discontinuity results from a drastic change in the topology of the energy landscape, which now exhibits one paramagnetic and two opposite ferromagnetic minima. Specifically, for $g<g_{dyn}$, the system is still in a dynamical ferromagnetic phase, where the orbits (green plots in Fig.\ref{fig:meanfieldresults}-(d)) are confined in the rightmost well and $\overline{\mathcal{S}}^x>0$. Then, approaching $g_{dyn}$ from below, $\overline{\mathcal{S}^x}$ still tends to the value determined by the position $\mathcal{S}^x_c$ of the nearest, non-vanishing local maximum. However, upon moving even slightly above $g_{dyn}$, the orbits (orange plots, Fig.\ref{fig:meanfieldresults}-(d)) become symmetric with respect to the $\mathbf{y}-\mathbf{z}$ plane, resulting in $\overline{\mathcal{S}^x}=0$ and marking a discontinuous dynamical transition.
From a broader perspective, the discontinuity arises from the existence of \emph{two maxima} of $\mathcal{H}_{cl}(\vec{\mathcal{S}},g_{dyn})$, having the same energy. These maxima are at located at $\mathcal{S}^x = \pm \mathcal{S}^x_c$, the two points where the separatrix (illustrated as a white trajectory in Fig.~\ref{fig:meanfieldresults}-(d)) has crossings.
As a consequence,  $\mathcal{S}^x(t)$ displays two distinct plateaus for $g\gtrsim g_{dyn}$, around $\mathcal{S}^x = \pm \mathcal{S}^x_c$ (see green and orange plots in Fig.~\ref{fig:meanfieldresults} (f)): when $g$ approaches $g_{dyn}$ from above, $\overline{\mathcal{S}^x}$ converges to the vanishing average of the two plateaus, creating a discontinuity at $g_{dyn}$.

\begin{figure*}[t!]
\centering
\includegraphics[width = \textwidth]{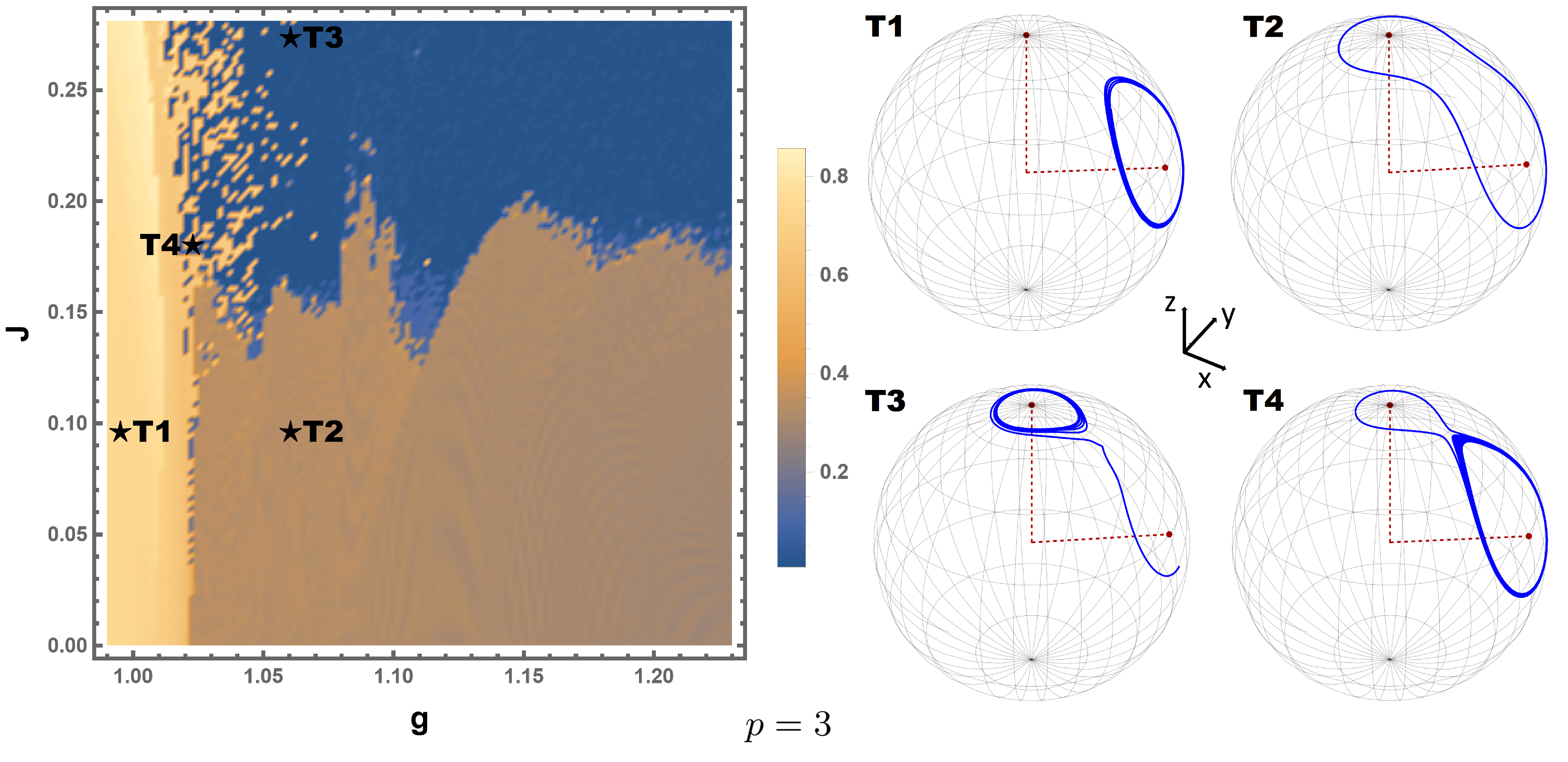}
\caption{(Color online) \textbf{(Left)} Non-equilibrium phase diagram for the time-average magnetization $\overline{\mathcal{S}^x}$, from Eq.~\eqref{eq:time_avg_magn}.  Each point of the phase diagram is obtained corresponds to a value of $\overline{\mathcal{S^x}}$, obtained by simultaneously Eqs.~\eqref{eq:dynamics_angles_NEQSWT} and~\eqref{spinwavesequations}, for a given choice of the couplings $g$ and $J$. The initial condition is fixed at $\Vec{\mathcal{S}}=\mathbf{x}$. The results shown in this figure refer to the case of $p=3$.
We fix the maximum simulation time as $T=2000$. Other simulation parameters are fixed at $N=100$ and $\lambda = 1$.
The color of each point on the phase diagram corresponds to a value of the time-averaged magnetization $\overline{\mathcal{S}^x}$, as specified in the interval shown on the right of the diagram. In particular, the yellow and the orange regions correspond to dynamical ferromagnetic and paramagnetic phases, respectively. 
\textbf{(Right)} The corresponding orbits are either confined in the ferromagnetic well (plot T1) or encircle both the two minima (plot T2). In the blue region, instead, is the dynamical paramagnetic region where the magnetization revolves around the minimum on the North pole (plot T3); here $\overline{\mathcal{S}^x}$ is closer to $0$ than in any other phase, though non-vanishing, and increases discontinuously when moving across the border with the orange zone. The crossover region is instead a narrow chaotic phase, reminiscent of the one found in \cite{Lerose2018}, where collective spin can localize at large times in either the paramagnetic or the ferromagnetic minima (plot T4).}
\label{fig:non-equilibrium phase diagrams 3}
\end{figure*}
\begin{figure*}[t]
\centering
\includegraphics[width = \textwidth]{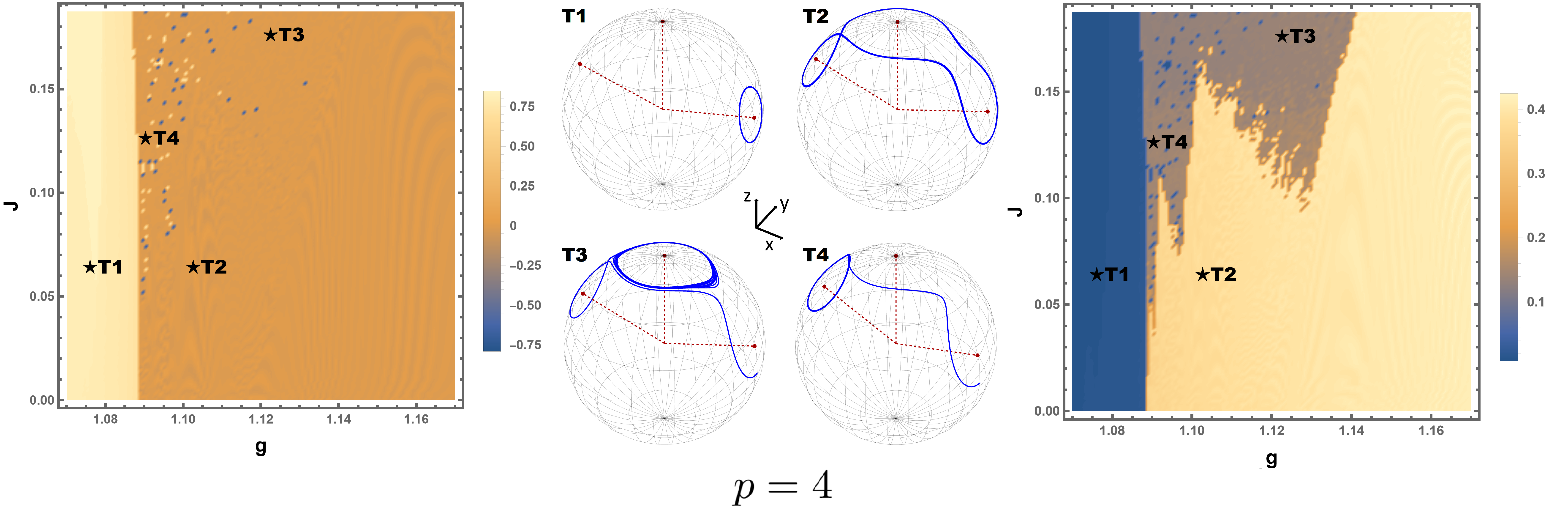}
\caption{(Color online) 
Simultaneous plot of the time-averaged magnetization $\overline{\mathcal{S}^x}$ (left) and time-averaged fluctuations $\overline{(\delta \mathcal{S}^x)^2}$ (right), for $p=4$. These are obtained integrating equations Eq.\eqref{eq:dynamics_angles_NEQSWT} and \eqref{spinwavesequations} for the same default parameters of $N$,$\lambda$ and the maximum time $T$ listed in the caption of Fig.\ref{fig:non-equilibrium phase diagrams 3}. For both the plots, each color corresponds to a value of the observable we plot, as specified in the legends reported on the right of each diagram.
\textbf{(Left)} From the plot of $\overline{\mathcal{S}^x}$, we identify two main regions: one corresponds to a dynamical ferromagnetic phase(yellow), where $\overline{\mathcal{S}^x}>0$ (plot T1), while the other is a paramagnetic phase (orange) where $\overline{\mathcal{S}^x}=0$. In between, some chaotic spots (blue and yellow spots) are found occasionally, where the magnetization eventually falls in one of the other two symmetric, ferromagnetic minima (plot T4).
\textbf{(Right)} Looking at the average fluctuations $\overline{(\delta \mathcal{S}^x)^2}$, we clearly see that the in dynamical paramagnetic phase, coinciding with the non-blue region of the phase diagram, can be split in a sub-phase 1 (yellow), where the dynamical trajectories either surround symmetrically all the three minima of
the landscape (plot T2) and a sub-phase 2 (blue), where the magnetization localizes (predominantly) in a paramagnetic well (plot T3). The border between the two identifies the transition line $J=J_{dyn}(g)$.}
\label{fig:non-equilibrium phase diagrams 4}
\end{figure*}
\section{Non-equilibrium spin-wave dynamics}

\subsection{Non-equilibrium spin-wave theory}

The dynamical transitions investigated in Sec.~\ref{SEC_mean-field_dynamics}-(b) originate from the coherent dynamics of the local spins in the $N\to\infty$ limit. In this limit, the spins are effectively decoupled from each other due to the all-to-all interaction among them. However, this coherence is expected to be unstable with respect to the inclusion of short-range interactions, which introduce fluctuations in the dynamics and drive the system towards eventual thermalization.
However, if the short-range coupling is small, the amplitude of these fluctuations is expected to be small for a parametrically long time, leaving possible instances of dynamical phases in the \emph{prethermal} stage of the dynamics~\cite{Mori2018preth,Gring2012prethermalization,Langen2016prethermalization}.\\

This scenario was investigated in Ref.~\cite{Lerose2018}, using the non-equilibrium spin-wave theory (NEQSWT). For $p=2$, it was shown that the dynamical critical point retrieved at the mean-field level is melted into an entire chaotic region of the phase diagram when short-range fluctuations are included.
In this work, we extend this analysis to the more general case of $p>2$, employing the NEQSWT to study the same quench dynamics discussed in Sec.~\ref{SEC_mean-field_dynamics}, under the influence of an extra, short-range term in the Hamiltonian. 
For the purposes of our study, we assume that our system is on a one-dimensional lattice with periodic boundary conditions, so that the short-range interaction term is expressed as~\cite{Lerose2018}:
\begin{equation}\label{eq:pert}
    \hat{U}= - J\sum_{i=1}^N \hat{\sigma}^x_i\hat{\sigma}^x_{i+1} + \frac JN \left(\sum_{i=1}^N \hat{\sigma}^x_i \right)^2 ~ .     
\end{equation}
To provide a clearer interpretation of Eq.~\eqref{eq:pert}, we introduce Fourier modes $\tilde{\sigma}^\alpha_{k}= \sum_{j=1}^N e^{-ikj} \hat{\sigma}_j^\alpha$, with $k = 2\pi n/N$ and $n =0, \dots, N-1$. Here, $N$ denotes the system size. Then, the full Hamiltonian is given by
\begin{equation}\label{eq:full_hamiltonian}
    \hat{H} = \hat{H}_0 +\hat{U} =- \frac{\lambda}{N^{p-1}} (\tilde{\sigma}^x_{0})^p - g \tilde{\sigma}^z_{0} - \frac{J}{N} \sum_{k \neq 0} \cos{k} ~ \tilde{\sigma}^x_k \tilde{\sigma}^x_{-k} ~ .
\end{equation}
$\hat{H}$ depends on the zero-momentum mode components $\Tilde{\sigma}_0^\alpha$, related to the magnetization through $\mathcal{S}^\alpha(t)=\braket{\Tilde{\sigma}_0^\alpha}/N$, solely via $\hat{H}_0$. On the other hand, the perturbation $\hat{U}$ from Eq.~\eqref{eq:pert} comprises only $k\neq 0$ contributions, which are the ones expected to induce dynamical fluctuations in the magnetization.
The expression in Eq.~\eqref{eq:pert} is then the simplest perturbation that breaks the permutation symmetry in Eq.~\eqref{eq:pspin_mf_hamiltonian}. Nevertheless, as we shall discuss later, our findings are expected to be independent of the range of interaction and of the dimensionality of the lattice. Hereafter, we outline the technical steps needed for implementing the non-equilibrium spin-wave theory (NEQSWT), for the Hamiltonian presented in Eq.~\eqref{eq:full_hamiltonian}.\\

The fundamental hypothesis underlying NEQSWT is that, at least as long as $J$ is sufficiently small, the net effect of the term from Eq.~\eqref{eq:pert} is to give rise to small spin-wave excitations on top of the classical magnetization $\vec{\mathcal{S}}(t)$. In particular, the magnetization length is still close to its maximal value, $|\vec{\mathcal{S}}(t)|\simeq 1$,  when the short-range fluctuations are sufficiently small. This allows the dynamics to be still effectively described by trajectories near the Bloch sphere, perturbed by fluctuations induced by the finite $k$ degrees of freedom.
The NEQSWT is then implemented by studying the dynamics in a time-dependent, rotating reference frame $\mathcal{R}$, identified by the time-dependent Cartesian vector basis $\{\mathbf{X}(t),\mathbf{Y}(t),\mathbf{Z}(t)\}$. The frame $\mathcal{R}$ is constructed such that the magnetization is aligned with the $\mathbf{Z}$-axis at any time $t$.
The base vectors of $\mathcal{R}$ can be identified by their components in the original frame $\{\mathbf{x},\mathbf{y},\mathbf{z}\}$, given by:
\begin{center}
$\mathbf{X}(t) = \begin{pmatrix}
    \cos{\theta(t)}\cos{\phi(t)} \\ \cos{\theta(t)}\sin{\phi(t)} \\ -\sin{\theta(t)}
\end{pmatrix} $ , \qquad 
$\mathbf{Y}(t) = \begin{pmatrix}
    -\sin{\phi(t)} \\ \cos{\phi(t)} \\ 0
\end{pmatrix} $ , \qquad 
$\mathbf{Z}(t) = \begin{pmatrix}
    \sin{\theta}\cos{\phi(t)} \\ \sin{\theta}\sin{\phi(t)} \\ \cos{\theta(t)}
\end{pmatrix} $ .
\end{center}
We implement this change of frame in the Hilbert space, through the time-dependent rotation operator $\hat{V}(t)=\exp\left(-i\phi(t) \sum_j \hat{\sigma}_j^z/2\right)\exp\left(-i\theta(t) \sum_j \hat{\sigma}_j^y/2\right)$, so that the spin operators transform accordingly:
\begin{equation}
\hat{V}(t) \hat{\sigma}_j^x \hat{V}^{\dagger}(t) = \mathbf{X}(t) \cdot \vec{\sigma}_j \equiv \hat{\sigma}_j^X(t), \qquad 
\hat{V}(t) \hat{\sigma}_j^y \hat{V}^{\dagger}(t) = \mathbf{Y}(t) \cdot \vec{\sigma}_j \equiv \hat{\sigma}_j^Y(t), \qquad 
\hat{V}(t) \hat{\sigma}_j^z \hat{V}^{\dagger}(t) = \mathbf{Z}(t) \cdot \vec{\sigma}_j \equiv \hat{\sigma}_j^Z(t) ~ .
\end{equation}
In the new frame $\mathcal{R}$, the Heisenberg equation of motion for the components $\hat{\sigma}_j^\alpha$, for $\alpha=\mathbf{X},\mathbf{Y},\mathbf{Z}$, can be written as
\begin{equation}
    i\hbar\frac{d}{dt}\hat{\sigma}_j^\alpha(t) =[\hat{\sigma}_j^\alpha(t), \Tilde{H}(t)] \ .
\end{equation}
Here, the modified Hamiltonian
\begin{equation}
    \Tilde{H}(t) = H + i \hat{V}(t)\partial_t \hat{V}^\dagger(t)
\end{equation}
includes an additional term $ i \hat{V}(t) \partial_t \hat{V}^\dagger(t) = -\vec{\omega}(t) \cdot \sum_j \vec{\sigma}_j(t)/2$, with $\vec{\omega}(t) = \big(-\sin{\theta}(t) \Dot{\phi}(t), -\Dot{\theta}(t) , \cos{\theta(t)}\Dot{\phi}(t) \big)$, plays a role analogous to the one of apparent forces in classical mechanics. 
The new Hamiltonian $\tilde{H}(t)$, describing the dynamics in the rotating frame, can be explictily written as:
\begin{equation}\label{eq:rotatingframeHamiltonian}
    \begin{split}
\frac{\widetilde{H}}{N} = &- g \left[ \left(\mathbf{X}\cdot\mathbf{z}\right) \frac{\tilde{\sigma}_{0}^X}{N} + \left(\mathbf{Y}\cdot\mathbf{z}\right) \frac{\tilde{\sigma}_{0}^Y}{N} +\left(\mathbf{Z}\cdot\mathbf{z}\right) \frac{\tilde{\sigma}_{0}^Z}{N}  \right]
  - \lambda\left[ \left(\mathbf{X}\cdot\mathbf{x}\right) \frac{\tilde{\sigma}_{0}^X}{N} + \left(\mathbf{Y}\cdot\mathbf{x}\right) \frac{\tilde{\sigma}_{0}^Y}{N} +\left(\mathbf{Z}\cdot\mathbf{x}\right) \frac{\tilde{\sigma}_{0}^Z}{N}  \right]^p \\
&- J \sum_{k\ne0} \cos k\left[ \left(\mathbf{X}\cdot\mathbf{x}\right) \frac{\tilde{\sigma}_{k}^X}{N} + \left(\mathbf{Y}\cdot\mathbf{x}\right) \frac{\tilde{\sigma}_{k}^Y}{N} +\left(\mathbf{Z}\cdot\mathbf{x}\right) \frac{\tilde{\sigma}_{k}^Z}{N}  \right]\left[ \left(\mathbf{X}\cdot\mathbf{x}\right) \frac{\tilde{\sigma}_{-k}^X}{N} + \left(\mathbf{Y}\cdot\mathbf{x}\right) \frac{\tilde{\sigma}_{-k}^Y}{N} +\left(\mathbf{Z}\cdot\mathbf{x}\right) \frac{\tilde{\sigma}_{-k}^Z}{N}  \right]   \\ 
&+\sin\theta \; s\dot{\phi} \; \frac{\tilde{\sigma}_0^X}{N} - s\dot{\theta} \; \frac{\tilde{\sigma}_0^Y}{N} - \cos\theta \; s\dot{\phi} \; \frac{\tilde{\sigma}_0^Z}{N} ~ .
\end{split}
\end{equation}
Here and in the following, we will often omit the time dependence of the operators and of the basis vectors to keep the notation compact.\\

$\Tilde{H}$ is the Hamiltonian describing the Heisenberg dynamics in the rotating frame $\mathcal{R}$, where the magnetization $\vec{\mathcal{S}}(t)$ is along $\mathbf{Z}(t)$. The time evolution of the angles $\theta(t)$ and $\phi(t)$ is determined by self-consistently
imposing that the transverse components of $\vec{\mathcal{S}}(t)$ vanish in the new frame, that is:
\begin{equation}\label{eq:self_cons_eqs}
    \mathcal{S}^X(t)=0 ~ , \qquad  \mathcal{S}^Y(t)=0 ~ .
\end{equation}
Solving Eqs.~\eqref{eq:self_cons_eqs} is in general a formidable task.
However, as long as the fluctuations transverse to the classical magnetization are small, we reasonably assume that the dynamics is dominated by terms at the lowest nontrivial order in the Holstein-Primakoff (HP) expansion~\cite{auerbach2012interacting}:
\begin{equation}
\label{eq:approxH-P}
\hat{\sigma}_i^X \sim  \frac{\hat{q}_i}{\sqrt{s}} ,\quad     \hat{\sigma}_i^Y  \sim \frac{\hat{p}_i}{\sqrt{s}}, \quad \hat{\sigma}_i^Z  = 1 - \frac{\hat{q}_i^2+\hat{p}_i^2-1}{2 s},
\end{equation}
where $s= 1/2$ for the current case.
As in the case of Ref. \cite{Lerose2018}, we retain perturbative terms from the HP expansion which are quadratic in the spin-wave modes $\Tilde{q}_k = \sum_j e^{-ikj}\hat{q}_j/\sqrt{N}$ and $\Tilde{p}_k =\sum_j e^{-ikj}\hat{p }_j/\sqrt{N}$, i.e. the Fourier transforms of $\hat{q}_i$ and $\hat{p}_i$.
This is equivalent to keep the following terms in $\Tilde{H}$:
\begin{gather}
    \begin{split}\label{eq:Hlin}
    \hat{U}_1 = &  \frac{\Tilde{q}_0}{\sqrt{Ns}}\Big\{s \sin{\theta}\Dot{\phi} - p \lambda (\mathbf{Z}\cdot\mathbf{x})^{p-1}(\mathbf{X}\cdot\mathbf{x})\Big(
    1 -(p-1)\frac{\hat{n}_{SW}}{Ns} \Big) - g (\mathbf{X}\cdot\mathbf{z}) + \\
    & +\frac{2(\mathbf{Z}\cdot\mathbf{x})}{Ns}\sum_{k \neq 0} J \cos k \big[(\mathbf{X}\cdot\mathbf{x})\Tilde{q}_k\Tilde{q}_{-k}+(\mathbf{Y}\cdot\mathbf{x})\frac{\Tilde{q}_k\Tilde{p}_{-k}+\Tilde{p}_k\Tilde{q}_{-k}}{2}\big]\Big\} +\\
    + & \frac{\Tilde{p}_0}{\sqrt{Ns}}\Big\{-s \Dot{\theta}-p \lambda (\mathbf{Z}\cdot\mathbf{x})^{p-1}(\mathbf{Y}\cdot\mathbf{x})\Big(
    1 -(p-1)\frac{\hat{n}_{SW}}{Ns} \Big) - g (\mathbf{Y}\cdot\mathbf{z}) + \\
    & +\frac{2(\mathbf{Z}\cdot\mathbf{x})}{Ns}\sum_{k \neq 0} J \cos k \big[(\mathbf{Y}\cdot\mathbf{x})\Tilde{p}_k\Tilde{p}_{-k}+(\mathbf{X}\cdot\mathbf{x})\frac{\Tilde{q}_k\Tilde{p}_{-k}+\Tilde{p}_k\Tilde{q}_{-k}}{2}\big]\Big\} ~ ,
    \end{split} \\
    \begin{split}\label{eq:U2_0_eq}
    \hat{U}^{(0)}_2 &= (p\lambda (\mathbf{Z}\cdot\mathbf{x})^p + g (\mathbf{Z}\cdot\mathbf{z}) + s \cos{\theta}\Dot{\phi}) \frac{1}{Ns}\sum_{k\neq 0}\hat{n}_k ~ ,
    \end{split}\\ 
    \begin{split}\label{eq:U2_eq}
    \hat{U}_2 &= -\frac{1}{Ns}\sum_{k\neq 0}J \cos k \{ (\mathbf{X}\cdot\mathbf{x})^2 \Tilde{q}_k\Tilde{q}_{-k} + (\mathbf{Y}\cdot\mathbf{x})^2 \Tilde{p}_k\Tilde{p}_{-k} +2(\mathbf{X}\cdot\mathbf{x})(\mathbf{Y}\cdot\mathbf{x})\frac{\Tilde{q}_k\Tilde{p}_{-k}+\Tilde{p}_k\Tilde{q}_{-k}}{2} \} ~ .
    \end{split}
\end{gather}
In Eq.~\eqref{eq:Hlin}, we also defined the spin-wave number operator $\hat{n}_{SW}=\sum_i (\hat{q}_i^2+\hat{p}_i^2)/2$.
Specifically, the terms $\hat{U}^{(0)}_2$ and $\hat{U}_2$ represent the quadratic terms from the HP expansion of the mean-field term $\hat{H}_0$ and the short-range perturbation $\hat{U}$, respectively.
Within this quadratic approximation, imposing the Eqs.~\eqref{eq:self_cons_eqs} is equivalent to requiring that the average of each coefficient, which appears either in front of $\Tilde{q}_0$ or $\Tilde{p}_0$ in Eq.~\eqref{eq:Hlin}, vanishes self-consistently. After some algebra, this requirement leads to the following equations of motion:
\begin{equation} \label{eq:dynamics_angles_NEQSWT}
    \left\{
    \begin{split}
    s\Dot{\phi} = & p \lambda (\sin{\theta})^{p-2}(\cos\phi)^p\cos\theta \big\{
    1 -(p-1)\epsilon(t) \big\} -g -2J \delta^{qq}(t) \cos\theta\cos^2\phi+ 2J \delta^{qp}(t) \sin{\phi}\cos{\phi}
    \\
    s \Dot{\theta} = & p \lambda (\sin{\theta}\cos{\phi})^{p-1}\sin{\phi}\{
    1 -(p-1)\epsilon(t)\} -2J\delta^{pp}(t) \sin{\theta}\sin{\phi}\cos{\phi} + 2J\delta^{qp}(t)\sin{\theta}\cos{\theta}\cos^2 \phi ~.
     \end{split}
    \right.
\end{equation}
In Eqs.~\eqref{eq:dynamics_angles_NEQSWT}, we defined the ``quantum feedback" variables
\begin{equation}\label{quantumfeedback}
\delta^{\alpha\beta}(t) \equiv  \frac{1}{Ns}\sum_{k\ne0} \Delta^{\alpha\beta}_\mathbf{k}(t) \cos k \ , 
\end{equation}
for $\alpha, \beta \in \{p,q\}$. 
These variables couple the classical spin to the corresponding spin-wave correlation functions, defined by
\begin{equation}\label{delta}
    \Delta^{qq}_k (t) \equiv  \left\langle \tilde{q}_k(t) \tilde{q}_{-k}(t) \right\rangle, \quad
    \Delta^{pp}_k (t) \equiv  \left\langle \tilde{p}_k(t) \tilde{p}_{-k}(t) \right\rangle, \quad
    \Delta^{qp}_k (t) \equiv \left\langle \tilde{q}_k(t) \tilde{p}_{-k}(t) + \tilde{p}_k(t) \tilde{q}_{-k}(t) \right\rangle/2 ~ ,
\end{equation}
defined for each value of $k = 2\pi n/N$, where $n=1,\dots,N-1$.
We also defined the spin-wave density,
\begin{equation} \label{eq:spinwave_density}
\epsilon(t) = \frac{1}{Ns}\sum_{k\neq 0} \frac{\Delta_k^{qq}(t)+\Delta_k^{pp}(t)-1}{2} ~ .
\end{equation}
The equations of motion for the spin-wave correlators are derived from the Heisenberg equations generated by the sum of the quadratic Hamiltonians $\hat{U}_2^{0}+\hat{U}_2$ and read as follows:
\begin{equation} \label{spinwavesequations}
\left\{
    \begin{split}
    s\frac{d}{dt}\Delta_k^{qq}&= (4 J \cos{k}\cos{\theta}\sin{\phi}\cos{\phi})\Delta_k^{qq}+ \{2p\lambda(\sin{\theta})^{p-2}(\cos{\phi})^p -4J \cos{k} \sin^2\phi\}\Delta_k^{qp} \\
    s\frac{d}{dt}\Delta_k^{qp} &= -\{p\lambda(\sin{\theta})^{p-2}(\cos{\phi})^p- 2 J \cos{k} \cos^2 \phi \cos^2\theta\}\Delta_k^{qq} +\{p\lambda(\sin{\theta})^{p-2}(\cos{\phi})^p -2J \cos{k} \sin^2\phi\}\Delta_k^{pp} \\
    s\frac{d}{dt}\Delta_k^{pp} &= -\{2p\lambda(\sin{\theta})^{p-2}(\cos{\phi})^p - 4 J \cos{k} \cos^2 \phi \cos^2\theta\}\Delta_k^{qp} -(4 J \cos{k}\cos{\theta}\sin{\phi}\cos{\phi})\Delta_k^{pp} ~ .
    \end{split}
\right.
\end{equation}
We refer the reader to Refs.~\cite{Lerose2018,Lerose2019} for a more detailed calculation.
We observe that, in the limit of $J \to 0$, the equations \eqref{eq:dynamics_angles_NEQSWT} decouple from the quantum feedback and consistently reduce to a representation of Eq.~\eqref{eq:Hamilton_equation} in the spherical coordinates $\theta(t)$ and $\phi(t)$.\\

The NEQSWT is expected to be valid as long as the density of spin-wave excitations is small, that is $\epsilon(t)\ll 1$.
In this case, the modulus of the magnetization $|\vec{\mathcal{S}}(t)|=1-\epsilon(t)$ is close to one and the dynamics can still be described in terms of classical trajectories.
In this regime, spin waves behave as free bosonic excitations, which interact with the
 macroscopic collective spin only, corresponding to the $k = 0$ mode. Higher-order terms appearing in Eq.~\eqref{eq:pspin_mf_hamiltonian}, which account for nonlinear scattering
among the spin waves, can be neglected: they are expected to contribute significantly to the dynamics only at longer times and to drive the system away from the prethermal regime relevant for the DPT discussed here~\cite{Lerose2019}.

\subsection{Modified non-equilibrium phase diagram}

The dynamics of the magnetization, as described by Eqs.~\eqref{eq:dynamics_angles_NEQSWT} , is still equivalent to the one of a particle moving into a multi-well shaped energy profile~\footnote{This is true as long as the $g$ is below the spinodal point $g_{sp}$ from Eq.~\eqref{eq:spinodal_point}, a condition always satisfied in our study.}, as discussed in Sec.~\ref{sec_sciolla_theory}. However, in this case, an additional damping effect arises from the exchange of energy with the spin-wave degrees of freedom, essentially acting as a self-generated bath. Further details on this mechanism are discussed in Sec.~\ref{SUBSEC_emission_mecanism}. The strength of the damping is controlled by the coupling $J$ in Eqs.~\eqref{eq:dynamics_angles_NEQSWT}. For $p=2$, this mechanism is responsible for the melting of the dynamical critical point into a chaotic crossover phase~\cite{Lerose2018,Lerose2019}.
In this phase, the magnetization asymptotically localizes into one of the two ferromagnetic wells, although the dynamics is strongly sensitive to both the initial conditions and the integration parameters. It is natural to ask if the spin-wave emission is going to have the same effect in the first-order case $p>2$. \\

\begin{figure}[t]
\includegraphics[width = .6\columnwidth]{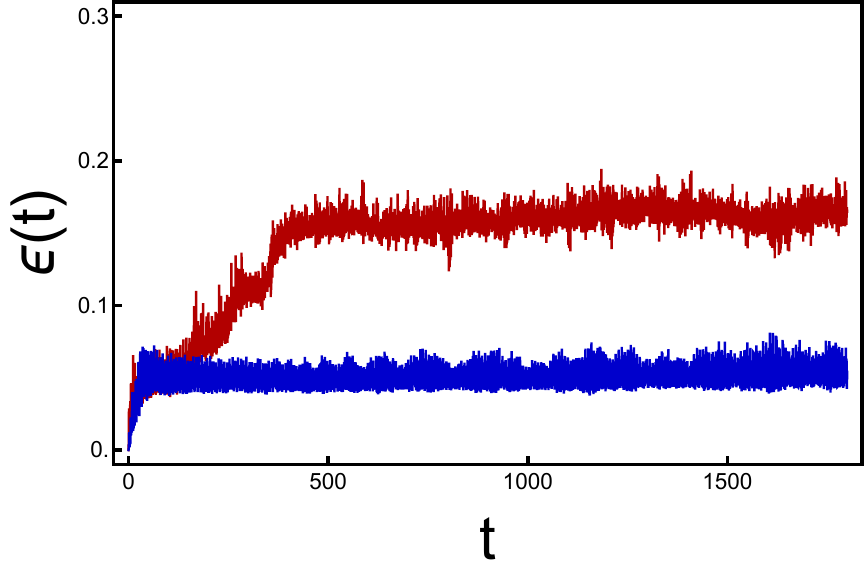}
\caption{(Color online) Plots of two examples of the profile of the function $\epsilon(t)$, defined in eq.~\eqref{eq:spinwave_density} and computed by within the NEQSW described in the main text, by fixing either $p=3$, $J=0.25$, $g=1.15$ (red plot) or $p=4$, $J=0.2$, $g=1.12$ (blue plot). In both plots we fix $N=100$ and $\lambda=1$.
}
\label{fig:spin_wave_density_fig}
\end{figure}

To understand the effect of fluctuations on the mean-field dynamical transition, we study the post-quench dynamics of the system at $J>0$, by  simultaneously integrating equations \eqref{eq:dynamics_angles_NEQSWT} and \eqref{spinwavesequations}, for several values of the couplings $g$ and $J$. 
Before the quench, we prepare again the system in the fully polarized state from Eq.~\eqref{eq:initial_state}, which is equivalent to the initial conditions $\phi(0)=0$ and $\theta(0)=\pi/2$. In this initial state, there are no spin-wave excitations.
For each choice of $g$ and $J$ and within our maximum simulation time, we verify the spin wave density $\epsilon(t)$ from Eq.~\eqref{eq:spinwave_density} always settles around a small
value, so that the dynamics is consistently in a prethermal regime.
In particular, as shown in Fig.~\ref{fig:spin_wave_density_fig}, $\epsilon(t)$ typically grows from zero and saturates to a small value. Consequently, the magnetization is asymptotically damped to a trajectory whose energy, $\mathcal{H}_{cl}(\vec{\mathcal{S}}(t),g)$, is slightly lower than the one at $t=0$. We define the new dynamical phases in terms of the topology of these trajectories, asymptotically reached after the damping. It is crucial to note once again that these new phases are \emph{prethermal}: they are expected to be disappear at longer times, as soon as the non-linear interaction among the spin-wave degrees of freedom is taken into account, leading to thermal relaxation of the system.
Below we will reconstruct the dynamical phase diagrams by looking at $\overline{\mathcal{S}^x}$ as a function of $g$ and $J$ and at individual trajectories. 
Similar to Sec.~\ref{SEC_mean-field_dynamics}, our focus will be on the cases of $p=3$ and $p=4$, which are paradigmatic for $p>2$ odd and $p>2$ even, respectively.\\

If spin waves were emitted at a constant rate in time, the orbits encompassing all the minima (i.e. orange trajectories in Fig.~\ref{fig:meanfieldresults}-(a) and (d) ) would localize down one of the wells of $\mathcal{H}_{cl}(\Vec{\mathcal{S}},g)$ with approximately equal probability. The phase diagram for $\overline{\mathcal{S}^x}$, depicted for $p=3$ in Fig.~\ref{fig:non-equilibrium phase diagrams 3}, reveals a more nuanced behavior. While the dynamical ferromagnetic phase is of course robust against fluctuations, the dynamical bistable phase loses stability for sufficiently large values of $J$ and its trajectories predominantly localize around the paramagnetic minimum, asymptotically converging to stable orbits. These asymptotic orbits identify a third, new \emph{dynamical paramagnetic phase} on the phase diagram.
Quantitatively, each dynamical phase corresponds to a narrow interval of values of $\overline{\mathcal{S}^x}$, with the greatest values in the dynamical ferromagnetic phase and the smallest in the paramagnetic one (although non-vanishing due to the asymmetry of the energy profile). 
In our discussion, we use the notation $J_{dyn}(g)$ to identify the transition line between the dynamical bistable and dynamical paramagnetic phases. As detailed in Appendix~\ref{APP_discontinuity_line}, $\overline{\mathcal{S}^x}$ exhibits a discontinuity when crossing $J_{dyn}(g)$, giving rise to a new first-order DPT driven by $J$.
The predominance of localization around $\mathcal{S}^x = 0$ is softened close to the mean-field critical point $g = g_{dyn}$, where localization in the ferromagnetic basin becomes more frequent. The line separating the dynamical ferromagnetic phase from other phases is consequently melted in a narrow chaotic crossover region, akin to the one observed for $p=2$~\cite{Lerose2018}.\\

As shown in the phase diagrams from Fig.~\ref{fig:non-equilibrium phase diagrams 4}, a similar phenomenon is observed for $p=4$. 
Here, as soon as $J$ is moved above a critical threshold $J_{dyn}(g)$ (generically distinct from the one retrieved for $p=3$), the trajectories from the mean-field dynamical paramagnetic phase, initially encompassing all the three minima of the landscape, become unstable and localize in the paramagnetic well. 
However, this discontinuity in the time-evolution can not be observed from $\overline{\mathcal{S}^x}$, displayed in Fig.~\ref{fig:non-equilibrium phase diagrams 4} (left). Specifically, $\overline{\mathcal{S}^x}$ is zero for orbits either surrounding all the minima or asymptotically localizing in the paramagnetic well, while $\overline{\mathcal{S}^x}>0$ in the ferromagnetic basin.
Thus, we also examine the behaviour of the time-averaged fluctuation~\cite{wang2021dissipative}, defined as
\begin{equation} \label{timefluctuations}
    \overline{(\delta \mathcal{S}^x)^2}= \lim_{T\to \infty} \frac{1}{T}\int_0^T dt \  \big(\mathcal{S}^x(t)-\overline{\mathcal{S}^x}\big)^2 \ .
\end{equation}
From the phase diagram in Fig.~\ref{fig:non-equilibrium phase diagrams 4} (right), it is clear that $\overline{(\delta \mathcal{S}^x)^2}$ is discontinuous across the transition line $J_{dyn}(g)$, where the topology of the asymptotic trajectories abruptly changes. As a consequence, the dynamical paramagnetic phase, corresponding to $\overline{\mathcal{S}^x}=0$, can be divided in two sub-phases:
\begin{itemize}
    \item[-] A \emph{dynamical paramagnetic phase} 1, where the asymptotic orbits surround all the minima, like in the corresponding mean-field phase.
    \item[-] A \emph{dynamical paramagnetic phase} 2, where the trajectories eventually localize in the paramagnetic well. 
\end{itemize}
Like in the case of $p=3$, these phases are identified by different narrow intervals of values of $\overline{(\delta\mathcal{S}^x)^2}$. The smaller values correspond to the dynamical paramagnetic phase 2 and to eventual localization in the paramagnetic basin.\\

\begin{figure*}
\centering
\includegraphics[width=1.05 \textwidth]{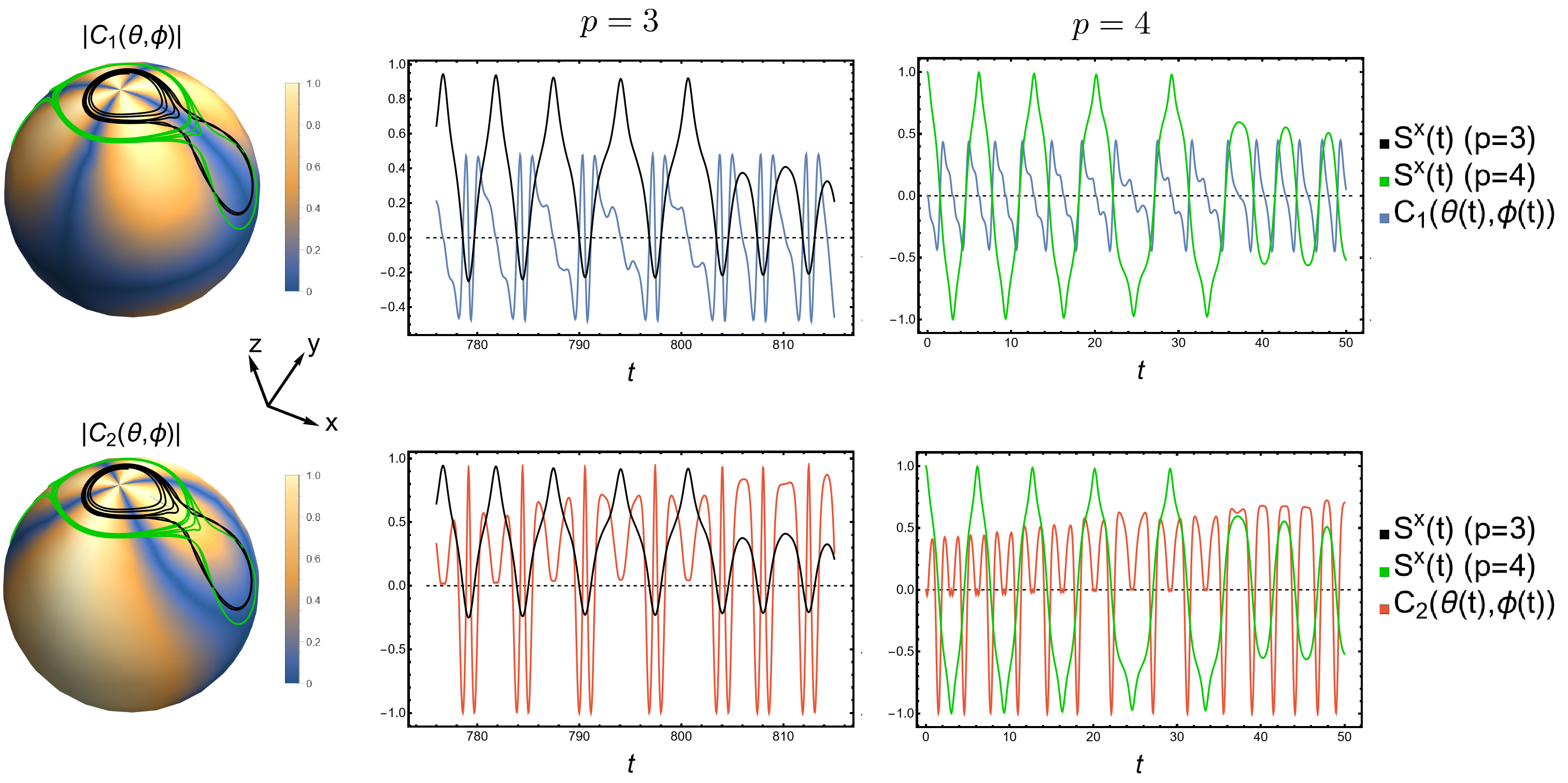}
\caption{(Color online)
Dynamics of two trajectories, parametrized by $\theta(t)$ and $\phi(t)$ and computed respectively for $p=3$, $g=1.09$, $J=0.26$ (black plots) and $p=4$, $g=1.13$, $J=0.15$ (green plots).
\textbf{(Left)} Spherical plots of the magnitude of the couplings $C_1(\theta,\phi)$ and $C_2(\theta,\phi)$, against the dynamical evolution of the two trajectories on the Bloch Sphere. 
\textbf{(Center)} Time evolution of the couplings $C_1(\theta(t),\phi(t))$ (blue) and $C_2(\theta(t),\phi(t))$ (red) along the trajectory obtained at $p=3$, each compared against the longitudinal magnetization $\mathcal{S}^x(t)=(1-\epsilon(t))\sin\phi(t) \cos\phi(t)$ (black).
\textbf{(Right)} Same plots represented at the center, this time for the trajectory computed at $p=4$. }
\label{fig:spinwaveemission}
\end{figure*} 

\subsection{The mechanism behind the localization of the magnetization}\label{SUBSEC_emission_mecanism}
Our results are very different from the ones found in Ref.~\cite{Lerose2018} for the case of $p=2$, where the same short-range perturbation from Eq.~\eqref{eq:pert} led to  dynamical chaos.
In this section we show that, despite the apparent difference, the dynamical phases retrieved both for $p=2$ and $p>2$ share a common origin, which can be related to the predominant emission of spin waves when the classical trajectory visits the paramagnetic well.\\

Intuitively, the inhomogeneous emission is due to the specific form of the short-range perturbation in Eq.~\eqref{eq:pert}, which induces fluctuations in the collective dynamics when the local spins are not aligned with the $x$-axis: as the maximal misalignment in the dynamics is reached when the magnetization crosses the plane $\mathcal{S}^x=0$,  spin waves are expected to be mostly emitted there.
The previous argument can be made quantitative by investigating the time evolution of the spin-wave density $\epsilon(t)$, which quantifies the degree of spin-wave excitations in the system. By differentiating both sides of Eq.~\eqref{eq:spinwave_density}, we obtain the spin-wave emission rate:
\begin{equation}
    \frac{d \epsilon(t)}{dt} = \frac{1}{2Ns} \sum_{k \neq 0} \big( \frac{d\Delta^{qq}_k(t)}{dt}+\frac{d\Delta^{pp}_k(t)}{dt}\big)
\end{equation}
Then, substituting Eqs.~\eqref{spinwavesequations} on the right-hand side we obtain the following evolution equation for $\epsilon(t)$:
\begin{equation}\label{eq:eom_swdensity}
\begin{split}
    \frac{d \epsilon(t)}{dt} &=-4\frac Js (\mathbf{X}(t)\cdot\mathbf{x})(\mathbf{Y}(t)\cdot\mathbf{x})~(\delta^{qq}(t)-\delta^{pp}(t) )+4\frac Js \big((\mathbf{X}(t)\cdot\mathbf{x})^2-(\mathbf{Y}(t)\cdot\mathbf{x})^2\big)\delta^{qp}(t)\\
    &=4\frac J s \cos\theta(t)\sin\phi(t)\cos\phi(t) ~(\delta^{qq}(t)-\delta^{pp}(t) ) + 4\frac J s  (\cos^2\theta(t) \cos^2\phi(t)-\sin^2\phi(t)) ~\delta^{qp}(t) ~ .
\end{split}
\end{equation}
The quantum feedback terms $\delta^{\alpha\beta}(t)$ are the ones defined in Eq.~\eqref{quantumfeedback}.\\

From Eq.~\eqref{eq:eom_swdensity}, it is evident that the emission rate $d\epsilon/dt$ depends explicitly on the position of the magnetization on the Bloch sphere. This dependency is encapsulated in the coefficients:
\begin{equation}\label{eq:coeffs}
    \begin{split}
    C_1(\theta,\phi)&=(\mathbf{X}\cdot\mathbf{x})(\mathbf{Y}\cdot\mathbf{x}) =\cos\theta\sin\phi\cos\phi \\
    C_2(\theta,\phi) &=(\mathbf{X}\cdot\mathbf{x})^2-(\mathbf{Y}\cdot\mathbf{x})^2 =\cos^2\theta \cos^2\phi-\sin^2\phi \end{split} ~ .
\end{equation}
We observe that the coefficients $C_1(\theta,\phi)$ and $C_2(\theta,\phi)$ do not depend on $p$. Instead, they are derived from a combination of the coefficients appearing in the Hamiltonian $\hat{U}_2$ from Eq.~\eqref{eq:U2_eq}, obtained by a quadratic expansion of the perturbation $\hat{U}$, from Eq.~\eqref{eq:pert}, in the moving frame $\mathcal{R}$. This observation aligns with the physical expectation that $\hat{U}$ is the only term in the Hamiltonian which generates spin-waves excitations. Both $C_1(\theta,\phi)$ and $C_2(\theta,\phi)$ vanish when the magnetization is along the $\mathbf{x}$-axis, i.e. $\theta=\pi/2$ and $\phi=0$.
In Fig.~\ref{fig:spinwaveemission}, we plot the time evolution of the amplitudes $|C_1(\theta,\phi)|$ and $|C_2(\theta,\phi)|$, along two sample trajectories from the dynamical paramagnetic phases retrieved for $p=3$ and $p=4$, respectively. The results shown therein confirm that these amplitudes are maximised when the magnetization crosses the plane $\mathcal{S}^x=0$, located in the paramagnetic well. The previous observations confirm our intuition that the maximum spin-wave emission occurs concomitantly with the maximal misalignment between $\vec{\mathcal{S}}(t)$ and $\mathbf{x}$.\\

As the predominant emission in the paramagnetic well is determined only by the short-range perturbation, the different phenomena observed for $p=2$ and $p>2$ respectively can be addressed to the different stability properties of paramagnetic the stationary point $\Vec{\mathcal{S}} = \mathbf{z}$ of the energy landscape in Eq.~\eqref{energylandscape}. For $p=2$, $\Vec{\mathcal{S}} = \mathbf{z}$ is unstable and symmetric fluctuations in the two wells~\footnote{We observe that the right-hand side of Eq.~\eqref{eq:eom_swdensity} is invariant under reflection $\phi \to \phi + \pi$ with respect to the $z$-axis, so that spin-wave emission are symmetric in the two wells for $p=2$.}
induce dynamical chaos~\cite{Lerose2018,Lerose2019}.
However, for $p>2$, $\Vec{\mathcal{S}} = \mathbf{z}$  becomes a stable minimum so that the magnetization moves on stationary orbits after being damped, giving birth to the dynamical paramagnetic regions shown in Fig. \ref{fig:non-equilibrium phase diagrams 3} (left) and Fig. \ref{fig:non-equilibrium phase diagrams 4} (right). This phenomenon is reminiscent of Hopf bifurcations~\cite{vulpiani2009chaos} occurring in classical dynamical systems.\\

It is also worth noticing that the mechanism by which the fluctuations induce the localization of the collective spin is slightly more subtle than a simple dissipation mechanism. In particular, we observe that when $\epsilon(t)>0$, the magnetization length is $|\vec{\mathcal{S}}(t)|=1-\epsilon(t)$ decreases, so that the dynamics in Eq.~\eqref{eq:dynamics_angles_NEQSWT} takes place in the time-dependent modified potential
\begin{equation}\label{renorm_potential}
    \begin{split}
    \mathcal{H}_{\epsilon(t)}(\theta,\phi) =& -(1-\epsilon(t))^p \{\lambda (\sin \theta \cos \phi)^p - \frac g{(1-\epsilon(t))^{p-1}} \sin \theta \} ~.
    \end{split}
\end{equation}
As shown in the animated plots (see Ancillary Files~\cite{animations}), the profile of the $\mathcal{H}_{\epsilon(t)}(\theta,\phi)$ is squeezed towards zero energy when $\epsilon(t)$ grows: this eventually leads the magnetization to be trapped in the paramagnetic region, where $\epsilon(t)$ exhibits large spikes, while the  $\epsilon(t)$ is nearly stationary across the ferromagnetic wells.\\

All the results presented in this section are expected to be independent of the range of interaction of the perturbation in Eq.~\eqref{eq:pert} and of the dimensionality of the lattice: replacing
the $k$-dependent couplings $J\cos(k)$ with generic $\tilde{J}_\mathbf{k}$~\footnote{Here $\mathbf{k}$ is a $d$-dimensional vector if the lattice has dimensionality $d>1$} leaves Eq.~\eqref{eq:eom_swdensity} invariant~\footnote{Up to replacing all the terms in the form of $J\delta^{\alpha\beta}(t)$ with the more generic expression $\delta^{\alpha\beta}(t) \equiv  \sum_{\mathbf{k}\ne\mathbf{0}} \Delta^{\alpha\beta}_k(t) \tilde{J}_\mathbf{k}/ (Ns)$.}
and spin-waves are still expected to be emitted around the plane $\mathcal{S}^x=0$, as explicitly shown for the case of $p=2$~\cite{Lerose2019}. On the other hand, if the short-range interaction was along a direction not coinciding with the $\mathbf{x}$-axis, the dependency of the coefficients from Eq.~\eqref{eq:coeffs} on the angles $\theta$ and $\phi$ could be different: in this case, spin-wave emission may occur in different regions on the Bloch sphere, leading to a different non-equilibrium phase diagram.

\section{Conclusions}
In conclusion, in this work we have studied the the post-quench dynamics of a fully-connected $p$-spin model (for $p>2$) perturbed by a short-ranged interaction, controlled by the coupling $J$, generalizing to arbitrary values of $p$ the system studied in previous work \cite{Sciolla2011,Lerose2018,Zunkovic18}. 
In the mean-field limit of $J=0$, the dynamics is equivalent to the one of a particle in a classical energy landscape. By identifying the topology of each orbit as a dynamical phase, we observe a dynamical phase transition (DPT) driven by the transverse field $g$. The order of the DPT depends on the parity of $p$, for $p>2$, which determines the qualitative shape of the effective potential. In particular, we find a second order dynamical transition for odd values of $p$ (where the potential is an asymmetric double-well) and first order one for an even $p$ (where the profile is made by three basins). 
The nature of this transition is modified by the presence of 
a short-range perturbation, treated in the framework of the non-equilibrium spin-wave theory. The latter
generates an effective damping in the dynamics which is maximal when the magnetization visits the paramagnetic well. The damping induces a prethermal stage of the dynamics and changes the nature of the dynamical phases, now being defined by the asymptotic behaviour of the magnetization. For $p\geq3$ odd, the short-range coupling drives a new first-order transition on the phase diagram,
while a more subtle transition appears for $p\geq 4$ even, detected in the order parameter fluctuations and being of the first-order like the one obtained in the mean-field limit.
%
Our analysis can be straightforwardly generalized to a wider class of fully-connected spin models with generic integrability breaking terms: the profile of the energy landscape and the axis where the integrability breaking interaction takes place are the only two ingredients which fully determine the features of the non-equilibrium phase diagram. \\

\emph{Acknowledgements }--- We thank A. Lerose and S. Pappalardi for interesting discussions and useful comments on the manuscript.

\begin{widetext}
\begin{appendix}

\section{The period of classical orbits and its relation with dynamical singularities}\label{APP_classical_period}

In this appendix we derive a closed formula for the period $T_{cl}(g)$ of the classical orbits discussed in Sec. \ref{SEC_mean-field_dynamics} of the main text. In particular, we show that $T_{cl}(g)$ diverges logarithmically when approaching the transition point $g=g_{dyn}$ both from above and below. We also show that the same singularity is retrieved also in $\overline{\mathcal{S}^x}(g)$, from Eq.~\eqref{eq:time_avg_magn}.\\

\begin{figure}[t]
\includegraphics[width = \textwidth]{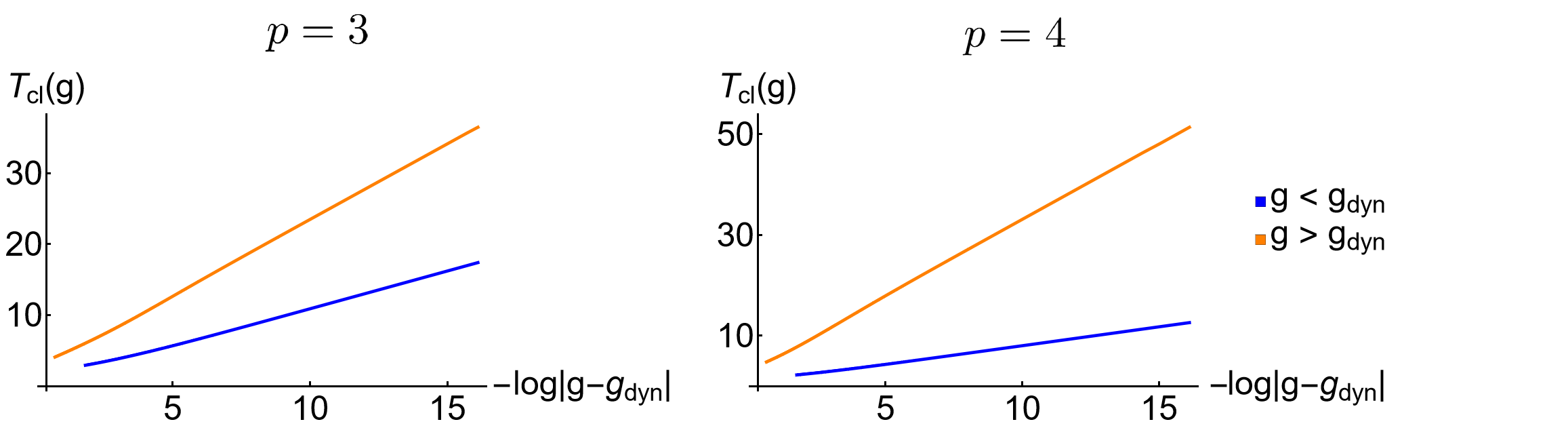}
\caption{Plot of the period $T_{cl}(g)$ of the classical trajectories originating from the dynamics discussed in Sec. \ref{SEC_mean-field_dynamics} of the main text, respectively for $p=3$ (left) and $p=4$ (right).
$T_{cl}(g)$ is compared against $\log |g-g_{dyn}|$, both for $g<g_{dyn}$ (blue) and $g>g_{dyn}$ (orange). $T_{cl}(g)$ is computed as outlined in Appendix~\ref{APP_classical_period}. We pose $\lambda=1$, like in the main text.}
\label{fig:periods}
\end{figure} 

To compute $T_{cl}(g)$, we first perform the change of variables:
\begin{equation}
\mathcal{S}^x = Q, \qquad
\mathcal{S}^y = \sqrt{1-Q^2} \sin(P), \qquad
\mathcal{S}^z = \sqrt{1-Q^2} \cos (P)
\end{equation}
where $Q \in [-1,1]$ and $P \in [0,\pi]$. By doing so, it can be shown~\cite{Sciolla2011,bapst2012quantum} that the dynamics described by the Eqs.~\eqref{eq:Hamilton_equation} is equivalent to an Hamilton dynamics, induced by the effective Hamiltonian
\begin{equation}\label{Hamiltonian_QP}
    \mathcal{H}_{cl}(Q,P) = -\lambda Q^P - g \sqrt{1-Q^2}\cos(P)
\end{equation}
and by the Poisson bracket $\{Q,P\}=1$.
The first of these new Hamilton equations is given by
\begin{equation}\label{Hamilton_eq_QP}
    \partial_t Q = g\sqrt{1-Q^2}\sin(P) ~ .
\end{equation}
Plugging eq. \eqref{Hamilton_eq_QP} into the expression \eqref{Hamiltonian_QP}, one straightforwardly obtains the following expression:
\begin{equation}
    (\mathcal{H}_{cl}(Q,P) + \lambda Q^p)^2 = g^2(1-Q^2)-(\partial_t Q)^2
\end{equation}
As the energy is conserved in the Hamilton dynamics, we can fix $\mathcal{H}_{cl}(Q(0),P(0)) = E_0 $,  where $E_0$ is the initial energy. Then, we can solve the dynamics by separation of variables to obtain
\begin{equation}\label{eq:sep_vars}
    t = \int_{\min(Q(0), Q(t)) }^{\max(Q(0), Q(t)) } \frac{dx}{\sqrt{ g^2 (1-x^2)-(E_0 + \lambda x^p)^2} }
\end{equation}
For the initial condition $\vec{\mathcal{S}}(0) = \mathbf{x}$ set in the main text, we have $E_0=-\lambda$, $Q(0) = 1$ and $P(0) = 0$. Plugging this information in Eq.~\eqref{eq:sep_vars}, we obtain a formula for the period of the classical orbits:
\begin{equation}\label{eq:APP_eq_classical_period}
    T_{cl}(g) = 2\int_{q_1}^1 \frac{dx}{\sqrt{g^2 (1-x^2)-(\lambda x^p - \lambda)^2} }
\end{equation}
where $q_1$ is the turning point of the orbit, obtained as the solution of the equation
\begin{equation}\label{eq:app_turning_point}
\mathcal{H}_{cl}(q_1,0)= -\lambda ~ ,
\end{equation} 
for $Q<1$.
We compute $T_{cl}(g)$ numerically from Eqs.~\eqref{eq:APP_eq_classical_period} and~\eqref{eq:app_turning_point}, for several values of $g$ and in the paradigmatic cases of $p=3$ and $p=4$.
The result, shown in Fig. \ref{fig:periods}, is that for $g\to g_{dyn}^\pm$ the period diverges as $T_{cl} \sim \log |g-g_{dyn}|^{-1}$, with a prefactor which is different above and below $g_{dyn}$.\\

The divergence of $T_{cl}(g)$ at $g_{dyn}$ is connected to change of topology of the underlying trajectories, being confined in a single ferromagnetic well for $g<g_{dyn}$ and exploring the whole landscape for $g>g_{dyn}$. In particular, at $g=g_{dyn}$ the trajectory is a \emph{separatrix}, that is a singular, non-periodic orbit. Any dynamics evolving on the separatrix asymptotically converges to the nearest local maximum $\vec{\mathcal{S}}_c$, separating the well where the motion takes place from the others. For any value of $p$, this implies that $\overline{\mathcal{S}^x}(g_{dyn}) = \mathcal{S}^x_c$.
Then, the divergence of
$T_{cl}(g)$ also generates a singularity in $\overline{\mathcal{S}^x}$, for $g\to g_{dyn}^-$.
In this case, the orbits evolving into the rightmost ferromagnetic well develop a plateau of diverging length near $\mathcal{S}^x_c$ (see Fig. \ref{fig:meanfieldresults}-(c) and (f) of the main text), so that $\overline{\mathcal{S}^x}$ can be estimated as
\begin{equation}\label{eq:APP_log_singularity_estimate}
\begin{split}
    \overline{\mathcal{S}^x}(g) = \frac 1{T_{cl}(g)}\int_0^{T_{cl}(g)}dt \mathcal{S}^x(t)
    = \mathcal{S}^x_c + \frac 1{T_{cl}(g)}\int_0^{T_{cl}(g)}dt \big(\mathcal{S}^x(t)-\mathcal{S}^x_c\big) \simeq \mathcal{S}^x_c + \frac c{T_{cl}(g)} ,
\end{split}
\end{equation}
Eq.~\eqref{eq:APP_log_singularity_estimate} holds under the reasonable assumption that the integral $\int_0^{T_{cl}(g)}dt \big(\mathcal{S}^x(t)-\mathcal{S}^x_c\big)$ is bounded and converges to a positive constant $c>0$ at the transition point.  Eq.~\eqref{eq:APP_log_singularity_estimate} implies 
that 
\begin{equation}
    \overline{\mathcal{S}^x}-\mathcal{S}^x_c \propto \frac 1{\log|g_{dyn}-g|^{-1}} ~ ,
\end{equation}
when approaching the transition from below.
The log-singularity is quantitatively confirmed by the results shown in Fig.~\ref{fig:mean-field_log_singularity}, obtained by computing $\overline{\mathcal{S}^x}$ numerically (as in Sec.\ref{SEC_mean-field_dynamics} of the main text) and $\mathcal{S}^x_c$ from Eqs.~\eqref{eq:pspin_clean_red_stat_eqs} and~\eqref{eq:pspin_clean_equate_energies} of the main text.
We remark that, while the log-singularity in the time-average magnetization $\overline{\mathcal{S}^x}$ are intimately connected to the one of the periods, its continuity is determined by the topology of the effective potential, as discussed in the main text.

\begin{figure}[t]
\includegraphics[width = .9 \textwidth]{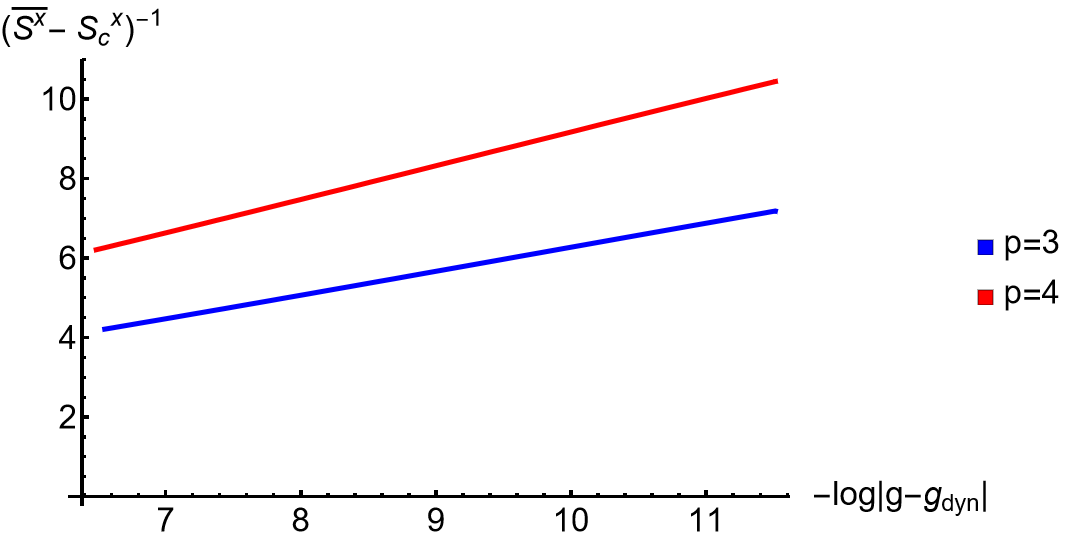}
\caption{Plot of the singularity of the time-average magnetization $\overline{\mathcal{S}^x}$, respectively for $p=3$ (blue) and $p=4$ (red). $\overline{\mathcal{S}^x}$ is plotted as a function of $g$ in the dynamical ferromagnetic phase, where $g<g_{dyn}$, and $\mathcal{S}^x_c$ is defined as an unstable stationary point of the potential in Eq.~\eqref{energylandscape} of the main text at $g_{dyn}$. We refer the reader to Appendix~\ref{APP_classical_period} for further details.}
\label{fig:mean-field_log_singularity}
\end{figure} 

\section{Details on the first-order transition line}\label{APP_discontinuity_line}

In this appendix, examine in greater detail the discontinuity line $J=J_{dyn}(g)$, appearing in Fig. \ref{fig:non-equilibrium phase diagrams 3} (left) and Fig.\ref{fig:non-equilibrium phase diagrams 4} (right) of the main text, analyzing the cases of $p=3$ and $p=4$ separately.\\

\begin{figure}[h]
\includegraphics[width = \textwidth]{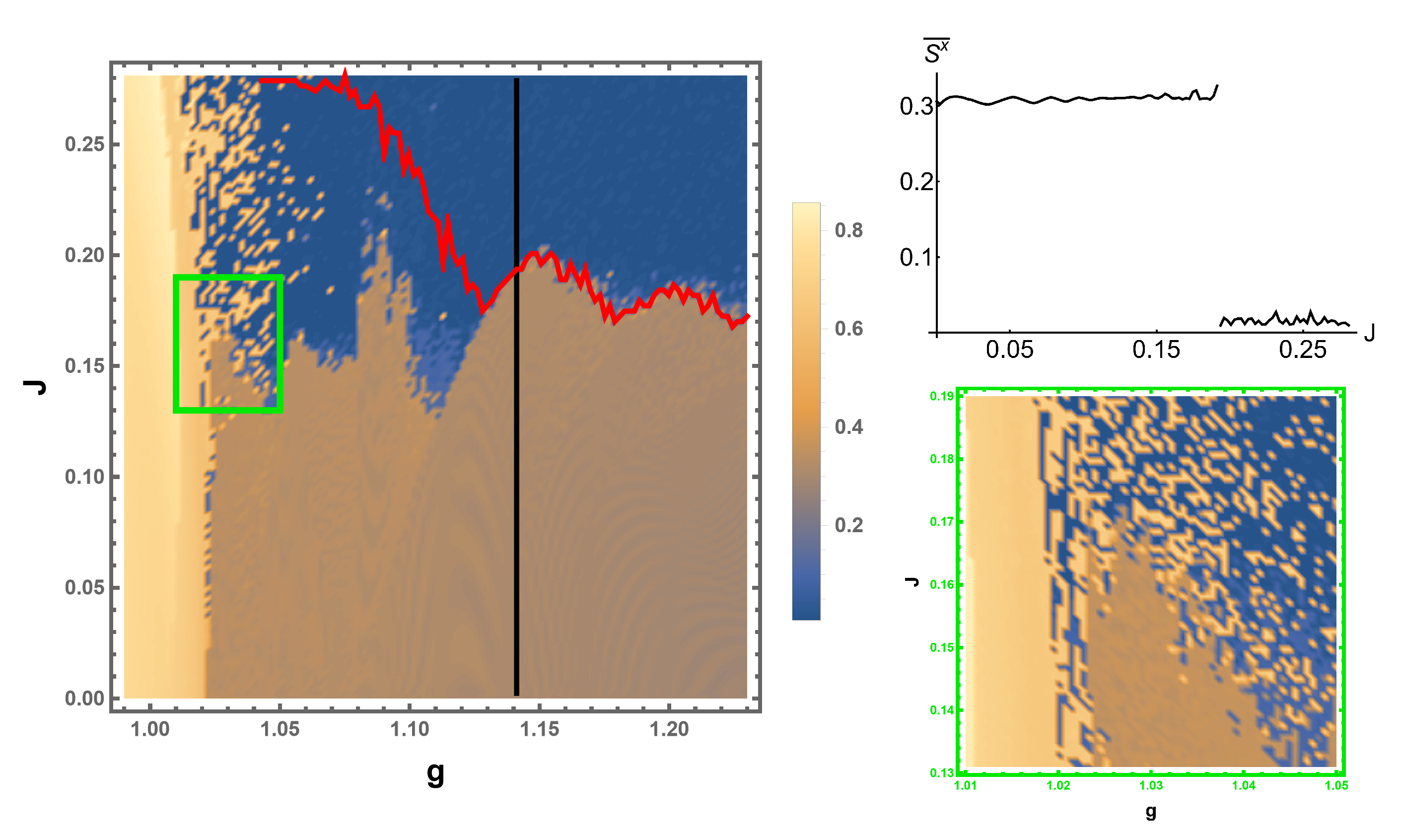}
\caption{{(Color online) Non-equilibrium phase diagram for the $p$-spin model in eq. \eqref{eq:pspin_mf_hamiltonian}, for $p=3$. \textbf{(Left)} Over the same phase diagram shown in Fig.~\ref{fig:non-equilibrium phase diagrams 3} (left) of the main text, we plot the threshold line from Eq.~\eqref{eq:app_threshold_sw_dens} (red), above which $\overline{\epsilon}>\epsilon_{sp}$ (see details in Appendix~\ref{APP_discontinuity_line}). The black vertical line and the green box indicate the values of $g$ and $J$ investigated in the plots on the right.
\textbf{(Top-right)} Plot of the time-averaged magnetization $\overline{\mathcal{S}^x}$, as function of $J$ and at fixed $g\simeq 1.142$ (black line on the phase diagram on the left). $\overline{\mathcal{S}^x}$ is discontinuous around $J \simeq 0.194$.
\textbf{(Bottom-right)} Inset from the phase diagram on the left (in the green box), around the chaotic region where the second-order and first-order transition line merge.}}
\label{fig:firstorder_transition_3.png}
\end{figure} 

For $p=3$, the transition line $J_{dyn}(g)$ emerges clearly fixing a sufficiently large value of $g$: the plot in Fig. \ref{fig:firstorder_transition_3.png}  (top-right) shows that the time-average magnetization $\overline{\mathcal{S}^x}$ has a discontinuity in $J$, though not being either completely smooth $J$ above and below the discontinuity point, because of the noise induced by the spin-wave emission. %
In this regime, it is possible to make a semi-analytical estimation of the point $J_{dyn}(g)$ where the transition happens, as discussed in the following. 
First we remind that, as discussed in Sec.~\ref{SUBSEC_emission_mecanism}, the effective potential determining the dynamics is squeezed by the emission of spin waves and is expressed by Eq.~\eqref{renorm_potential} of the main text.
From this expression, it is evident that the dynamics is driven by an effective transverse field
\begin{equation}~\label{eq:app_threshold_sw_dens}
    g_{\epsilon} = \frac{g}{(1-\epsilon)^{p-1}} ~ .
\end{equation}
Here, $g_{\epsilon}$ depends on time through the spin-wave density $\epsilon(t)$, defined in Eq.~\eqref{eq:spinwave_density}.
However, Eq.~\eqref{eq:spinodal_point} also indicates the existence of a critical value of $g$, denoted as $g_{sp}$, such that for $g>g_{sp}$ the energy landscape becomes a single well.
These observations leads us to the conclusion that it exist a threshold value $\epsilon_{sp}$ for the spin-wave density, defined by the equation
\begin{equation}
    \frac{g}{(1-\epsilon)^{p-1}} = g_{sp} ~ ,
\end{equation}
such that the squeezed potential from Eq.~\eqref{renorm_potential} displays a single paramagnetic well whenever $\epsilon>\epsilon_{sp}$.
Then, a \emph{sufficient} condition for the localization of the magnetization in the paramagnetic well corresponds to $\epsilon(t)$ asymptotically exceeding $\epsilon_{sp}$. Quantitatively, this corresponds to:
\begin{equation}\label{eq:app_suff_cond_loc}
    \overline{\epsilon} = \lim_{T\to \infty} \frac 1T \int_0^T\epsilon(t) > \epsilon_{sp}
\end{equation}
Eq.~\eqref{eq:app_suff_cond_loc} condition implies that, for large times, the squeezed potential from Eq.~\eqref{renorm_potential} displays a single, paramagnetic well, where the magnetization is by definition localized.
The plots in Fig.\ref{fig:firstorder_transition_3.png} (left) show that, fixing a sufficiently large $g$, the values of $J$ where $\overline{\epsilon}$ overcomes the threshold $\epsilon_{sp}$ matches the transition point $J_{dyn}(g)$: this provides a semi-analytical argument for the prediction of $J_{dyn}(g)$ at large $g$, as the threshold $\epsilon_{sp}$ can be predicted analytically, but there is no explicit formula relating $J_{dyn}(g)$ to $\overline{\epsilon}$.
Our argument fails for smaller $g$, where the localization mechanism becomes more subtle (as discussed in Sec.\ref{SUBSEC_emission_mecanism} the main text) and the first-order transition line is smeared into a chaotic crossover close to the mean-field critical point $g_{dyn}$. A similar fate happens to the second-order critical line, extending from the point $(g,J)=(g_{dyn},0)$, to finite values of $J$, so that the chaotic crossover prevents the possibility of a precise estimation of the tricritical point where the two lines meet. However, we can roughly identify the transition point as the center of the
finite-width area where the two transition lines merge completely with the chaotic region: in the inset in Fig. \ref{fig:firstorder_transition_3.png} (bottom-right), we show that this happens approximately
around $(g, J) \simeq (1.026, 0.17)$.\\
For $p=4$, the results in Fig.
\ref{fig:firstorder_transition_4.png}
(right) show a discontinuous transition driven by $J$, this time detected by the time-averaged fluctuations $\overline{(\delta\mathcal{S}^x)^2}$, even though the semi-analytical estimation of $J_{dyn}(g)$ fails in this case. Moreover, here both the mean-field dynamical transition, driven by $g$, as well as the one driven by $J$, are of the first-order, so that we do not retrieve the tricritical behaviour discussed for $p=3$.

\begin{figure}[h]
\includegraphics[width = \textwidth]{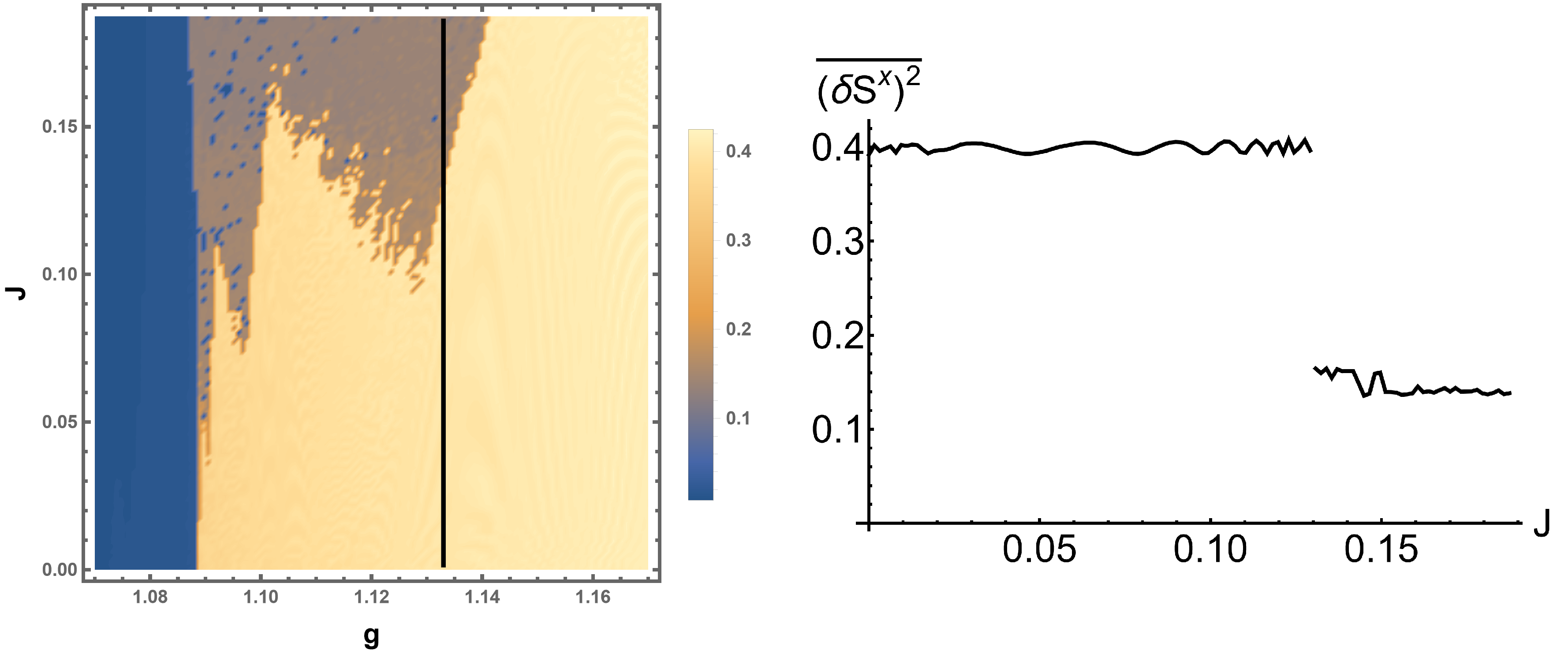}
\caption{(Color online) Non-equilibrium phase diagram for the $p$-spin model in eq. \eqref{eq:pspin_mf_hamiltonian}, for $p=4$. \textbf{(Left)} Same phase diagram shown in Fig.~\ref{fig:non-equilibrium phase diagrams 4} (left). The black vertical line and the green box indicate the values of $g$ and $J$ investigated in the plots on the right.
\textbf{(Right)} Plot of the time-averaged fluctuations $\overline{(\delta\mathcal{S}^x)^2}$, as function of $J$ and at fixed $g\simeq 1.133$ (black line on the phase diagram on the left). $\overline{(\delta\mathcal{S}^x)^2}$ is discontinuous around $J \simeq 0.131$.
}
\label{fig:firstorder_transition_4.png}
\end{figure} 

\end{appendix}
\end{widetext}
\bibliography{bibliography.bib}

\begin{thebibliography}{65}%
\makeatletter
\providecommand \@ifxundefined [1]{%
 \@ifx{#1\undefined}
}%
\providecommand \@ifnum [1]{%
 \ifnum #1\expandafter \@firstoftwo
 \else \expandafter \@secondoftwo
 \fi
}%
\providecommand \@ifx [1]{%
 \ifx #1\expandafter \@firstoftwo
 \else \expandafter \@secondoftwo
 \fi
}%
\providecommand \natexlab [1]{#1}%
\providecommand \enquote  [1]{``#1''}%
\providecommand \bibnamefont  [1]{#1}%
\providecommand \bibfnamefont [1]{#1}%
\providecommand \citenamefont [1]{#1}%
\providecommand \href@noop [0]{\@secondoftwo}%
\providecommand \href [0]{\begingroup \@sanitize@url \@href}%
\providecommand \@href[1]{\@@startlink{#1}\@@href}%
\providecommand \@@href[1]{\endgroup#1\@@endlink}%
\providecommand \@sanitize@url [0]{\catcode `\\12\catcode `\$12\catcode `\&12\catcode `\#12\catcode `\^12\catcode `\_12\catcode `\%12\relax}%
\providecommand \@@startlink[1]{}%
\providecommand \@@endlink[0]{}%
\providecommand \url  [0]{\begingroup\@sanitize@url \@url }%
\providecommand \@url [1]{\endgroup\@href {#1}{\urlprefix }}%
\providecommand \urlprefix  [0]{URL }%
\providecommand \Eprint [0]{\href }%
\providecommand \doibase [0]{http://dx.doi.org/}%
\providecommand \selectlanguage [0]{\@gobble}%
\providecommand \bibinfo  [0]{\@secondoftwo}%
\providecommand \bibfield  [0]{\@secondoftwo}%
\providecommand \translation [1]{[#1]}%
\providecommand \BibitemOpen [0]{}%
\providecommand \bibitemStop [0]{}%
\providecommand \bibitemNoStop [0]{.\EOS\space}%
\providecommand \EOS [0]{\spacefactor3000\relax}%
\providecommand \BibitemShut  [1]{\csname bibitem#1\endcsname}%
\let\auto@bib@innerbib\@empty
\bibitem [{\citenamefont {Polkovnikov}\ \emph {et~al.}(2011)\citenamefont {Polkovnikov}, \citenamefont {Sengupta}, \citenamefont {Silva},\ and\ \citenamefont {Vengalattore}}]{Polkovnikov2011}%
  \BibitemOpen
  \bibfield  {author} {\bibinfo {author} {\bibfnamefont {A.}~\bibnamefont {Polkovnikov}}, \bibinfo {author} {\bibfnamefont {K.}~\bibnamefont {Sengupta}}, \bibinfo {author} {\bibfnamefont {A.}~\bibnamefont {Silva}}, \ and\ \bibinfo {author} {\bibfnamefont {M.}~\bibnamefont {Vengalattore}},\ }\href {\doibase 10.1103/revmodphys.83.863} {\bibfield  {journal} {\bibinfo  {journal} {Reviews of Modern Physics}\ }\textbf {\bibinfo {volume} {83}},\ \bibinfo {pages} {863} (\bibinfo {year} {2011})}\BibitemShut {NoStop}%
\bibitem [{\citenamefont {Rossini}\ \emph {et~al.}(2009)\citenamefont {Rossini}, \citenamefont {Silva}, \citenamefont {Mussardo},\ and\ \citenamefont {Santoro}}]{Rossini09}%
  \BibitemOpen
  \bibfield  {author} {\bibinfo {author} {\bibfnamefont {D.}~\bibnamefont {Rossini}}, \bibinfo {author} {\bibfnamefont {A.}~\bibnamefont {Silva}}, \bibinfo {author} {\bibfnamefont {G.}~\bibnamefont {Mussardo}}, \ and\ \bibinfo {author} {\bibfnamefont {G.~E.}\ \bibnamefont {Santoro}},\ }\href {\doibase 10.1103/PhysRevLett.102.127204} {\bibfield  {journal} {\bibinfo  {journal} {Phys. Rev. Lett.}\ }\textbf {\bibinfo {volume} {102}},\ \bibinfo {pages} {127204} (\bibinfo {year} {2009})}\BibitemShut {NoStop}%
\bibitem [{\citenamefont {Calabrese}\ \emph {et~al.}(2011)\citenamefont {Calabrese}, \citenamefont {Essler},\ and\ \citenamefont {Fagotti}}]{Calabrese11}%
  \BibitemOpen
  \bibfield  {author} {\bibinfo {author} {\bibfnamefont {P.}~\bibnamefont {Calabrese}}, \bibinfo {author} {\bibfnamefont {F.~H.~L.}\ \bibnamefont {Essler}}, \ and\ \bibinfo {author} {\bibfnamefont {M.}~\bibnamefont {Fagotti}},\ }\href {\doibase 10.1103/PhysRevLett.106.227203} {\bibfield  {journal} {\bibinfo  {journal} {Phys. Rev. Lett.}\ }\textbf {\bibinfo {volume} {106}},\ \bibinfo {pages} {227203} (\bibinfo {year} {2011})}\BibitemShut {NoStop}%
\bibitem [{\citenamefont {Maraga}\ \emph {et~al.}(2014)\citenamefont {Maraga}, \citenamefont {Smacchia}, \citenamefont {Fabrizio},\ and\ \citenamefont {Silva}}]{Maraga14}%
  \BibitemOpen
  \bibfield  {author} {\bibinfo {author} {\bibfnamefont {A.}~\bibnamefont {Maraga}}, \bibinfo {author} {\bibfnamefont {P.}~\bibnamefont {Smacchia}}, \bibinfo {author} {\bibfnamefont {M.}~\bibnamefont {Fabrizio}}, \ and\ \bibinfo {author} {\bibfnamefont {A.}~\bibnamefont {Silva}},\ }\href {\doibase 10.1103/PhysRevB.90.041111} {\bibfield  {journal} {\bibinfo  {journal} {Phys. Rev. B}\ }\textbf {\bibinfo {volume} {90}},\ \bibinfo {pages} {041111} (\bibinfo {year} {2014})}\BibitemShut {NoStop}%
\bibitem [{\citenamefont {Polkovnikov}(2005)}]{Polkovnikov05}%
  \BibitemOpen
  \bibfield  {author} {\bibinfo {author} {\bibfnamefont {A.}~\bibnamefont {Polkovnikov}},\ }\href {\doibase 10.1103/PhysRevB.72.161201} {\bibfield  {journal} {\bibinfo  {journal} {Phys. Rev. B}\ }\textbf {\bibinfo {volume} {72}},\ \bibinfo {pages} {161201} (\bibinfo {year} {2005})}\BibitemShut {NoStop}%
\bibitem [{\citenamefont {Zurek}\ \emph {et~al.}(2005)\citenamefont {Zurek}, \citenamefont {Dorner},\ and\ \citenamefont {Zoller}}]{Zurek05}%
  \BibitemOpen
  \bibfield  {author} {\bibinfo {author} {\bibfnamefont {W.~H.}\ \bibnamefont {Zurek}}, \bibinfo {author} {\bibfnamefont {U.}~\bibnamefont {Dorner}}, \ and\ \bibinfo {author} {\bibfnamefont {P.}~\bibnamefont {Zoller}},\ }\href {\doibase 10.1103/PhysRevLett.95.105701} {\bibfield  {journal} {\bibinfo  {journal} {Phys. Rev. Lett.}\ }\textbf {\bibinfo {volume} {95}},\ \bibinfo {pages} {105701} (\bibinfo {year} {2005})}\BibitemShut {NoStop}%
\bibitem [{\citenamefont {Dziarmaga}(2005)}]{Dziarmaga05}%
  \BibitemOpen
  \bibfield  {author} {\bibinfo {author} {\bibfnamefont {J.}~\bibnamefont {Dziarmaga}},\ }\href {\doibase 10.1103/PhysRevLett.95.245701} {\bibfield  {journal} {\bibinfo  {journal} {Phys. Rev. Lett.}\ }\textbf {\bibinfo {volume} {95}},\ \bibinfo {pages} {245701} (\bibinfo {year} {2005})}\BibitemShut {NoStop}%
\bibitem [{\citenamefont {Cherng}\ and\ \citenamefont {Levitov}(2006)}]{Cherng06}%
  \BibitemOpen
  \bibfield  {author} {\bibinfo {author} {\bibfnamefont {R.~W.}\ \bibnamefont {Cherng}}\ and\ \bibinfo {author} {\bibfnamefont {L.~S.}\ \bibnamefont {Levitov}},\ }\href {\doibase 10.1103/PhysRevA.73.043614} {\bibfield  {journal} {\bibinfo  {journal} {Phys. Rev. A}\ }\textbf {\bibinfo {volume} {73}},\ \bibinfo {pages} {043614} (\bibinfo {year} {2006})}\BibitemShut {NoStop}%
\bibitem [{\citenamefont {King}\ \emph {et~al.}(2022)\citenamefont {King}, \citenamefont {Suzuki}, \citenamefont {Raymond}, \citenamefont {Zucca}, \citenamefont {Lanting}, \citenamefont {Altomare}, \citenamefont {Berkley}, \citenamefont {Ejtemaee}, \citenamefont {Hoskinson}, \citenamefont {Huang}, \citenamefont {Ladizinsky}, \citenamefont {MacDonald}, \citenamefont {Marsden}, \citenamefont {Oh}, \citenamefont {Poulin-Lamarre}, \citenamefont {Reis}, \citenamefont {Rich}, \citenamefont {Sato}, \citenamefont {Whittaker}, \citenamefont {Yao}, \citenamefont {Harris}, \citenamefont {Lidar}, \citenamefont {Nishimori},\ and\ \citenamefont {Amin}}]{King2022JZ}%
  \BibitemOpen
  \bibfield  {author} {\bibinfo {author} {\bibfnamefont {A.~D.}\ \bibnamefont {King}}, \bibinfo {author} {\bibfnamefont {S.}~\bibnamefont {Suzuki}}, \bibinfo {author} {\bibfnamefont {J.}~\bibnamefont {Raymond}}, \bibinfo {author} {\bibfnamefont {A.}~\bibnamefont {Zucca}}, \bibinfo {author} {\bibfnamefont {T.}~\bibnamefont {Lanting}}, \bibinfo {author} {\bibfnamefont {F.}~\bibnamefont {Altomare}}, \bibinfo {author} {\bibfnamefont {A.~J.}\ \bibnamefont {Berkley}}, \bibinfo {author} {\bibfnamefont {S.}~\bibnamefont {Ejtemaee}}, \bibinfo {author} {\bibfnamefont {E.}~\bibnamefont {Hoskinson}}, \bibinfo {author} {\bibfnamefont {S.}~\bibnamefont {Huang}}, \bibinfo {author} {\bibfnamefont {E.}~\bibnamefont {Ladizinsky}}, \bibinfo {author} {\bibfnamefont {A.~J.~R.}\ \bibnamefont {MacDonald}}, \bibinfo {author} {\bibfnamefont {G.}~\bibnamefont {Marsden}}, \bibinfo {author} {\bibfnamefont {T.}~\bibnamefont {Oh}}, \bibinfo {author} {\bibfnamefont {G.}~\bibnamefont {Poulin-Lamarre}}, \bibinfo {author} {\bibfnamefont
  {M.}~\bibnamefont {Reis}}, \bibinfo {author} {\bibfnamefont {C.}~\bibnamefont {Rich}}, \bibinfo {author} {\bibfnamefont {Y.}~\bibnamefont {Sato}}, \bibinfo {author} {\bibfnamefont {J.~D.}\ \bibnamefont {Whittaker}}, \bibinfo {author} {\bibfnamefont {J.}~\bibnamefont {Yao}}, \bibinfo {author} {\bibfnamefont {R.}~\bibnamefont {Harris}}, \bibinfo {author} {\bibfnamefont {D.~A.}\ \bibnamefont {Lidar}}, \bibinfo {author} {\bibfnamefont {H.}~\bibnamefont {Nishimori}}, \ and\ \bibinfo {author} {\bibfnamefont {M.~H.}\ \bibnamefont {Amin}},\ }\href {\doibase 10.1038/s41567-022-01741-6} {\bibfield  {journal} {\bibinfo  {journal} {Nature Physics}\ }\textbf {\bibinfo {volume} {18}},\ \bibinfo {pages} {1324–1328} (\bibinfo {year} {2022})}\BibitemShut {NoStop}%
\bibitem [{\citenamefont {Du}\ \emph {et~al.}(2023)\citenamefont {Du}, \citenamefont {Fang}, \citenamefont {Won}, \citenamefont {De}, \citenamefont {Huang}, \citenamefont {Xu}, \citenamefont {You}, \citenamefont {Gómez-Ruiz}, \citenamefont {del Campo},\ and\ \citenamefont {Cheong}}]{Du2023KZ}%
  \BibitemOpen
  \bibfield  {author} {\bibinfo {author} {\bibfnamefont {K.}~\bibnamefont {Du}}, \bibinfo {author} {\bibfnamefont {X.}~\bibnamefont {Fang}}, \bibinfo {author} {\bibfnamefont {C.}~\bibnamefont {Won}}, \bibinfo {author} {\bibfnamefont {C.}~\bibnamefont {De}}, \bibinfo {author} {\bibfnamefont {F.-T.}\ \bibnamefont {Huang}}, \bibinfo {author} {\bibfnamefont {W.}~\bibnamefont {Xu}}, \bibinfo {author} {\bibfnamefont {H.}~\bibnamefont {You}}, \bibinfo {author} {\bibfnamefont {F.~J.}\ \bibnamefont {Gómez-Ruiz}}, \bibinfo {author} {\bibfnamefont {A.}~\bibnamefont {del Campo}}, \ and\ \bibinfo {author} {\bibfnamefont {S.-W.}\ \bibnamefont {Cheong}},\ }\href {\doibase 10.1038/s41567-023-02112-5} {\bibfield  {journal} {\bibinfo  {journal} {Nature Physics}\ }\textbf {\bibinfo {volume} {19}},\ \bibinfo {pages} {1495–1501} (\bibinfo {year} {2023})}\BibitemShut {NoStop}%
\bibitem [{\citenamefont {Bando}\ \emph {et~al.}(2020)\citenamefont {Bando}, \citenamefont {Susa}, \citenamefont {Oshiyama}, \citenamefont {Shibata}, \citenamefont {Ohzeki}, \citenamefont {G\'omez-Ruiz}, \citenamefont {Lidar}, \citenamefont {Suzuki}, \citenamefont {del Campo},\ and\ \citenamefont {Nishimori}}]{Bando2023KZ}%
  \BibitemOpen
  \bibfield  {author} {\bibinfo {author} {\bibfnamefont {Y.}~\bibnamefont {Bando}}, \bibinfo {author} {\bibfnamefont {Y.}~\bibnamefont {Susa}}, \bibinfo {author} {\bibfnamefont {H.}~\bibnamefont {Oshiyama}}, \bibinfo {author} {\bibfnamefont {N.}~\bibnamefont {Shibata}}, \bibinfo {author} {\bibfnamefont {M.}~\bibnamefont {Ohzeki}}, \bibinfo {author} {\bibfnamefont {F.~J.}\ \bibnamefont {G\'omez-Ruiz}}, \bibinfo {author} {\bibfnamefont {D.~A.}\ \bibnamefont {Lidar}}, \bibinfo {author} {\bibfnamefont {S.}~\bibnamefont {Suzuki}}, \bibinfo {author} {\bibfnamefont {A.}~\bibnamefont {del Campo}}, \ and\ \bibinfo {author} {\bibfnamefont {H.}~\bibnamefont {Nishimori}},\ }\href {\doibase 10.1103/PhysRevResearch.2.033369} {\bibfield  {journal} {\bibinfo  {journal} {Phys. Rev. Res.}\ }\textbf {\bibinfo {volume} {2}},\ \bibinfo {pages} {033369} (\bibinfo {year} {2020})}\BibitemShut {NoStop}%
\bibitem [{\citenamefont {Braun}\ \emph {et~al.}(2015)\citenamefont {Braun}, \citenamefont {Friesdorf}, \citenamefont {Hodgman}, \citenamefont {Schreiber}, \citenamefont {Ronzheimer}, \citenamefont {Riera}, \citenamefont {del Rey}, \citenamefont {Bloch}, \citenamefont {Eisert},\ and\ \citenamefont {Schneider}}]{Braun15}%
  \BibitemOpen
  \bibfield  {author} {\bibinfo {author} {\bibfnamefont {S.}~\bibnamefont {Braun}}, \bibinfo {author} {\bibfnamefont {M.}~\bibnamefont {Friesdorf}}, \bibinfo {author} {\bibfnamefont {S.~S.}\ \bibnamefont {Hodgman}}, \bibinfo {author} {\bibfnamefont {M.}~\bibnamefont {Schreiber}}, \bibinfo {author} {\bibfnamefont {J.~P.}\ \bibnamefont {Ronzheimer}}, \bibinfo {author} {\bibfnamefont {A.}~\bibnamefont {Riera}}, \bibinfo {author} {\bibfnamefont {M.}~\bibnamefont {del Rey}}, \bibinfo {author} {\bibfnamefont {I.}~\bibnamefont {Bloch}}, \bibinfo {author} {\bibfnamefont {J.}~\bibnamefont {Eisert}}, \ and\ \bibinfo {author} {\bibfnamefont {U.}~\bibnamefont {Schneider}},\ }\href {\doibase 10.1073/pnas.1408861112} {\bibfield  {journal} {\bibinfo  {journal} {Proceedings of the National Academy of Sciences}\ }\textbf {\bibinfo {volume} {112}},\ \bibinfo {pages} {3641} (\bibinfo {year} {2015})},\ \Eprint {http://arxiv.org/abs/https://www.pnas.org/content/112/12/3641.full.pdf}
  {https://www.pnas.org/content/112/12/3641.full.pdf} \BibitemShut {NoStop}%
\bibitem [{\citenamefont {Anquez}\ \emph {et~al.}(2016)\citenamefont {Anquez}, \citenamefont {Robbins}, \citenamefont {Bharath}, \citenamefont {Boguslawski}, \citenamefont {Hoang},\ and\ \citenamefont {Chapman}}]{Anquez16}%
  \BibitemOpen
  \bibfield  {author} {\bibinfo {author} {\bibfnamefont {M.}~\bibnamefont {Anquez}}, \bibinfo {author} {\bibfnamefont {B.~A.}\ \bibnamefont {Robbins}}, \bibinfo {author} {\bibfnamefont {H.~M.}\ \bibnamefont {Bharath}}, \bibinfo {author} {\bibfnamefont {M.}~\bibnamefont {Boguslawski}}, \bibinfo {author} {\bibfnamefont {T.~M.}\ \bibnamefont {Hoang}}, \ and\ \bibinfo {author} {\bibfnamefont {M.~S.}\ \bibnamefont {Chapman}},\ }\href {\doibase 10.1103/PhysRevLett.116.155301} {\bibfield  {journal} {\bibinfo  {journal} {Phys. Rev. Lett.}\ }\textbf {\bibinfo {volume} {116}},\ \bibinfo {pages} {155301} (\bibinfo {year} {2016})}\BibitemShut {NoStop}%
\bibitem [{\citenamefont {Clark}\ \emph {et~al.}(2016)\citenamefont {Clark}, \citenamefont {Feng},\ and\ \citenamefont {Chin}}]{Clark16}%
  \BibitemOpen
  \bibfield  {author} {\bibinfo {author} {\bibfnamefont {L.~W.}\ \bibnamefont {Clark}}, \bibinfo {author} {\bibfnamefont {L.}~\bibnamefont {Feng}}, \ and\ \bibinfo {author} {\bibfnamefont {C.}~\bibnamefont {Chin}},\ }\href {\doibase 10.1126/science.aaf9657} {\bibfield  {journal} {\bibinfo  {journal} {Science}\ }\textbf {\bibinfo {volume} {354}},\ \bibinfo {pages} {606} (\bibinfo {year} {2016})},\ \Eprint {http://arxiv.org/abs/https://science.sciencemag.org/content/354/6312/606.full.pdf} {https://science.sciencemag.org/content/354/6312/606.full.pdf} \BibitemShut {NoStop}%
\bibitem [{\citenamefont {Anderson}\ \emph {et~al.}(2017)\citenamefont {Anderson}, \citenamefont {Clark}, \citenamefont {Crawford}, \citenamefont {Glatz}, \citenamefont {Aranson}, \citenamefont {Scherpelz}, \citenamefont {Feng}, \citenamefont {Chin},\ and\ \citenamefont {Levin}}]{Clark17}%
  \BibitemOpen
  \bibfield  {author} {\bibinfo {author} {\bibfnamefont {B.~M.}\ \bibnamefont {Anderson}}, \bibinfo {author} {\bibfnamefont {L.~W.}\ \bibnamefont {Clark}}, \bibinfo {author} {\bibfnamefont {J.}~\bibnamefont {Crawford}}, \bibinfo {author} {\bibfnamefont {A.}~\bibnamefont {Glatz}}, \bibinfo {author} {\bibfnamefont {I.~S.}\ \bibnamefont {Aranson}}, \bibinfo {author} {\bibfnamefont {P.}~\bibnamefont {Scherpelz}}, \bibinfo {author} {\bibfnamefont {L.}~\bibnamefont {Feng}}, \bibinfo {author} {\bibfnamefont {C.}~\bibnamefont {Chin}}, \ and\ \bibinfo {author} {\bibfnamefont {K.}~\bibnamefont {Levin}},\ }\href {\doibase 10.1103/PhysRevLett.118.220401} {\bibfield  {journal} {\bibinfo  {journal} {Phys. Rev. Lett.}\ }\textbf {\bibinfo {volume} {118}},\ \bibinfo {pages} {220401} (\bibinfo {year} {2017})}\BibitemShut {NoStop}%
\bibitem [{\citenamefont {Kang}\ \emph {et~al.}(2017)\citenamefont {Kang}, \citenamefont {Seo}, \citenamefont {Kim},\ and\ \citenamefont {Shin}}]{Kang17}%
  \BibitemOpen
  \bibfield  {author} {\bibinfo {author} {\bibfnamefont {S.}~\bibnamefont {Kang}}, \bibinfo {author} {\bibfnamefont {S.~W.}\ \bibnamefont {Seo}}, \bibinfo {author} {\bibfnamefont {J.~H.}\ \bibnamefont {Kim}}, \ and\ \bibinfo {author} {\bibfnamefont {Y.}~\bibnamefont {Shin}},\ }\href {\doibase 10.1103/PhysRevA.95.053638} {\bibfield  {journal} {\bibinfo  {journal} {Phys. Rev. A}\ }\textbf {\bibinfo {volume} {95}},\ \bibinfo {pages} {053638} (\bibinfo {year} {2017})}\BibitemShut {NoStop}%
\bibitem [{\citenamefont {Keesling}\ \emph {et~al.}(2019)\citenamefont {Keesling}, \citenamefont {Omran}, \citenamefont {Levine}, \citenamefont {Bernien}, \citenamefont {Pichler}, \citenamefont {Choi}, \citenamefont {Samajdar}, \citenamefont {Schwartz}, \citenamefont {Silvi}, \citenamefont {Sachdev}, \citenamefont {Zoller}, \citenamefont {Endres}, \citenamefont {Greiner}, \citenamefont {Vuleti{\'c}},\ and\ \citenamefont {Lukin}}]{Keesling19}%
  \BibitemOpen
  \bibfield  {author} {\bibinfo {author} {\bibfnamefont {A.}~\bibnamefont {Keesling}}, \bibinfo {author} {\bibfnamefont {A.}~\bibnamefont {Omran}}, \bibinfo {author} {\bibfnamefont {H.}~\bibnamefont {Levine}}, \bibinfo {author} {\bibfnamefont {H.}~\bibnamefont {Bernien}}, \bibinfo {author} {\bibfnamefont {H.}~\bibnamefont {Pichler}}, \bibinfo {author} {\bibfnamefont {S.}~\bibnamefont {Choi}}, \bibinfo {author} {\bibfnamefont {R.}~\bibnamefont {Samajdar}}, \bibinfo {author} {\bibfnamefont {S.}~\bibnamefont {Schwartz}}, \bibinfo {author} {\bibfnamefont {P.}~\bibnamefont {Silvi}}, \bibinfo {author} {\bibfnamefont {S.}~\bibnamefont {Sachdev}}, \bibinfo {author} {\bibfnamefont {P.}~\bibnamefont {Zoller}}, \bibinfo {author} {\bibfnamefont {M.}~\bibnamefont {Endres}}, \bibinfo {author} {\bibfnamefont {M.}~\bibnamefont {Greiner}}, \bibinfo {author} {\bibfnamefont {V.}~\bibnamefont {Vuleti{\'c}}}, \ and\ \bibinfo {author} {\bibfnamefont {M.~D.}\ \bibnamefont {Lukin}},\ }\href {\doibase 10.1038/s41586-019-1070-1}
  {\bibfield  {journal} {\bibinfo  {journal} {Nature}\ }\textbf {\bibinfo {volume} {568}},\ \bibinfo {pages} {207} (\bibinfo {year} {2019})}\BibitemShut {NoStop}%
\bibitem [{\citenamefont {Cui}\ \emph {et~al.}(2020)\citenamefont {Cui}, \citenamefont {G{\'o}mez-Ruiz}, \citenamefont {Huang}, \citenamefont {Li}, \citenamefont {Guo},\ and\ \citenamefont {del Campo}}]{Cui20}%
  \BibitemOpen
  \bibfield  {author} {\bibinfo {author} {\bibfnamefont {J.-M.}\ \bibnamefont {Cui}}, \bibinfo {author} {\bibfnamefont {F.~J.}\ \bibnamefont {G{\'o}mez-Ruiz}}, \bibinfo {author} {\bibfnamefont {Y.-F.}\ \bibnamefont {Huang}}, \bibinfo {author} {\bibfnamefont {C.-F.}\ \bibnamefont {Li}}, \bibinfo {author} {\bibfnamefont {G.-C.}\ \bibnamefont {Guo}}, \ and\ \bibinfo {author} {\bibfnamefont {A.}~\bibnamefont {del Campo}},\ }\href {\doibase 10.1038/s42005-020-0306-6} {\bibfield  {journal} {\bibinfo  {journal} {Communications Physics}\ }\textbf {\bibinfo {volume} {3}},\ \bibinfo {pages} {44} (\bibinfo {year} {2020})}\BibitemShut {NoStop}%
\bibitem [{\citenamefont {\ifmmode~\acute{S}\else \'{S}\fi{}wis\l{}ocki}\ \emph {et~al.}(2013)\citenamefont {\ifmmode~\acute{S}\else \'{S}\fi{}wis\l{}ocki}, \citenamefont {Witkowska}, \citenamefont {Dziarmaga},\ and\ \citenamefont {Matuszewski}}]{Swislocki13}%
  \BibitemOpen
  \bibfield  {author} {\bibinfo {author} {\bibfnamefont {T.}~\bibnamefont {\ifmmode~\acute{S}\else \'{S}\fi{}wis\l{}ocki}}, \bibinfo {author} {\bibfnamefont {E.}~\bibnamefont {Witkowska}}, \bibinfo {author} {\bibfnamefont {J.}~\bibnamefont {Dziarmaga}}, \ and\ \bibinfo {author} {\bibfnamefont {M.}~\bibnamefont {Matuszewski}},\ }\href {\doibase 10.1103/PhysRevLett.110.045303} {\bibfield  {journal} {\bibinfo  {journal} {Phys. Rev. Lett.}\ }\textbf {\bibinfo {volume} {110}},\ \bibinfo {pages} {045303} (\bibinfo {year} {2013})}\BibitemShut {NoStop}%
\bibitem [{\citenamefont {Coulamy}\ \emph {et~al.}(2017)\citenamefont {Coulamy}, \citenamefont {Saguia},\ and\ \citenamefont {Sarandy}}]{Coulamy17}%
  \BibitemOpen
  \bibfield  {author} {\bibinfo {author} {\bibfnamefont {I.~B.}\ \bibnamefont {Coulamy}}, \bibinfo {author} {\bibfnamefont {A.}~\bibnamefont {Saguia}}, \ and\ \bibinfo {author} {\bibfnamefont {M.~S.}\ \bibnamefont {Sarandy}},\ }\href {\doibase 10.1103/PhysRevE.95.022127} {\bibfield  {journal} {\bibinfo  {journal} {Phys. Rev. E}\ }\textbf {\bibinfo {volume} {95}},\ \bibinfo {pages} {022127} (\bibinfo {year} {2017})}\BibitemShut {NoStop}%
\bibitem [{\citenamefont {Shimizu}\ \emph {et~al.}(2018)\citenamefont {Shimizu}, \citenamefont {Hirano}, \citenamefont {Park}, \citenamefont {Kuno},\ and\ \citenamefont {Ichinose}}]{Shimizu18}%
  \BibitemOpen
  \bibfield  {author} {\bibinfo {author} {\bibfnamefont {K.}~\bibnamefont {Shimizu}}, \bibinfo {author} {\bibfnamefont {T.}~\bibnamefont {Hirano}}, \bibinfo {author} {\bibfnamefont {J.}~\bibnamefont {Park}}, \bibinfo {author} {\bibfnamefont {Y.}~\bibnamefont {Kuno}}, \ and\ \bibinfo {author} {\bibfnamefont {I.}~\bibnamefont {Ichinose}},\ }\href {\doibase 10.1103/PhysRevA.98.063603} {\bibfield  {journal} {\bibinfo  {journal} {Phys. Rev. A}\ }\textbf {\bibinfo {volume} {98}},\ \bibinfo {pages} {063603} (\bibinfo {year} {2018})}\BibitemShut {NoStop}%
\bibitem [{\citenamefont {Del~Re}\ \emph {et~al.}(2016)\citenamefont {Del~Re}, \citenamefont {Fabrizio},\ and\ \citenamefont {Tosatti}}]{DelRe16}%
  \BibitemOpen
  \bibfield  {author} {\bibinfo {author} {\bibfnamefont {L.}~\bibnamefont {Del~Re}}, \bibinfo {author} {\bibfnamefont {M.}~\bibnamefont {Fabrizio}}, \ and\ \bibinfo {author} {\bibfnamefont {E.}~\bibnamefont {Tosatti}},\ }\href {\doibase 10.1103/PhysRevB.93.125131} {\bibfield  {journal} {\bibinfo  {journal} {Phys. Rev. B}\ }\textbf {\bibinfo {volume} {93}},\ \bibinfo {pages} {125131} (\bibinfo {year} {2016})}\BibitemShut {NoStop}%
\bibitem [{\citenamefont {Sinha}\ \emph {et~al.}(2021)\citenamefont {Sinha}, \citenamefont {Chanda},\ and\ \citenamefont {Dziarmaga}}]{Sinha21}%
  \BibitemOpen
  \bibfield  {author} {\bibinfo {author} {\bibfnamefont {A.}~\bibnamefont {Sinha}}, \bibinfo {author} {\bibfnamefont {T.}~\bibnamefont {Chanda}}, \ and\ \bibinfo {author} {\bibfnamefont {J.}~\bibnamefont {Dziarmaga}},\ }\href {\doibase 10.1103/PhysRevB.103.L220302} {\bibfield  {journal} {\bibinfo  {journal} {Phys. Rev. B}\ }\textbf {\bibinfo {volume} {103}},\ \bibinfo {pages} {L220302} (\bibinfo {year} {2021})}\BibitemShut {NoStop}%
\bibitem [{\citenamefont {Lagnese}\ \emph {et~al.}(2021)\citenamefont {Lagnese}, \citenamefont {Surace}, \citenamefont {Kormos},\ and\ \citenamefont {Calabrese}}]{lagnese21}%
  \BibitemOpen
  \bibfield  {author} {\bibinfo {author} {\bibfnamefont {G.}~\bibnamefont {Lagnese}}, \bibinfo {author} {\bibfnamefont {F.~M.}\ \bibnamefont {Surace}}, \bibinfo {author} {\bibfnamefont {M.}~\bibnamefont {Kormos}}, \ and\ \bibinfo {author} {\bibfnamefont {P.}~\bibnamefont {Calabrese}},\ }\href@noop {} {\enquote {\bibinfo {title} {False vacuum decay in quantum spin chains},}\ } (\bibinfo {year} {2021}),\ \Eprint {http://arxiv.org/abs/2107.10176} {arXiv:2107.10176 [cond-mat.stat-mech]} \BibitemShut {NoStop}%
\bibitem [{\citenamefont {Qiu}\ \emph {et~al.}(2020)\citenamefont {Qiu}, \citenamefont {Liang}, \citenamefont {Yang}, \citenamefont {Yang}, \citenamefont {Tian}, \citenamefont {Xu},\ and\ \citenamefont {Duan}}]{Qiu21}%
  \BibitemOpen
  \bibfield  {author} {\bibinfo {author} {\bibfnamefont {L.-Y.}\ \bibnamefont {Qiu}}, \bibinfo {author} {\bibfnamefont {H.-Y.}\ \bibnamefont {Liang}}, \bibinfo {author} {\bibfnamefont {Y.-B.}\ \bibnamefont {Yang}}, \bibinfo {author} {\bibfnamefont {H.-X.}\ \bibnamefont {Yang}}, \bibinfo {author} {\bibfnamefont {T.}~\bibnamefont {Tian}}, \bibinfo {author} {\bibfnamefont {Y.}~\bibnamefont {Xu}}, \ and\ \bibinfo {author} {\bibfnamefont {L.-M.}\ \bibnamefont {Duan}},\ }\href {\doibase 10.1126/sciadv.aba7292} {\bibfield  {journal} {\bibinfo  {journal} {Science Advances}\ }\textbf {\bibinfo {volume} {6}} (\bibinfo {year} {2020}),\ 10.1126/sciadv.aba7292},\ \Eprint {http://arxiv.org/abs/https://advances.sciencemag.org/content/6/21/eaba7292.full.pdf} {https://advances.sciencemag.org/content/6/21/eaba7292.full.pdf} \BibitemShut {NoStop}%
\bibitem [{\citenamefont {Heyl}\ \emph {et~al.}(2013)\citenamefont {Heyl}, \citenamefont {Polkovnikov},\ and\ \citenamefont {Kehrein}}]{heyl2013dynamical}%
  \BibitemOpen
  \bibfield  {author} {\bibinfo {author} {\bibfnamefont {M.}~\bibnamefont {Heyl}}, \bibinfo {author} {\bibfnamefont {A.}~\bibnamefont {Polkovnikov}}, \ and\ \bibinfo {author} {\bibfnamefont {S.}~\bibnamefont {Kehrein}},\ }\href@noop {} {\bibfield  {journal} {\bibinfo  {journal} {Phys. Rev. Lett.}\ }\textbf {\bibinfo {volume} {110}},\ \bibinfo {pages} {135704} (\bibinfo {year} {2013})}\BibitemShut {NoStop}%
\bibitem [{\citenamefont {Heyl}(2018)}]{Heyl2018}%
  \BibitemOpen
  \bibfield  {author} {\bibinfo {author} {\bibfnamefont {M.}~\bibnamefont {Heyl}},\ }\href {\doibase 10.1088/1361-6633/aaaf9a} {\bibfield  {journal} {\bibinfo  {journal} {Reports on Progress in Physics}\ }\textbf {\bibinfo {volume} {81}},\ \bibinfo {pages} {054001} (\bibinfo {year} {2018})}\BibitemShut {NoStop}%
\bibitem [{\citenamefont {Weidinger}\ \emph {et~al.}(2017)\citenamefont {Weidinger}, \citenamefont {Heyl}, \citenamefont {Silva},\ and\ \citenamefont {Knap}}]{Weidinger17}%
  \BibitemOpen
  \bibfield  {author} {\bibinfo {author} {\bibfnamefont {S.~A.}\ \bibnamefont {Weidinger}}, \bibinfo {author} {\bibfnamefont {M.}~\bibnamefont {Heyl}}, \bibinfo {author} {\bibfnamefont {A.}~\bibnamefont {Silva}}, \ and\ \bibinfo {author} {\bibfnamefont {M.}~\bibnamefont {Knap}},\ }\href {\doibase 10.1103/PhysRevB.96.134313} {\bibfield  {journal} {\bibinfo  {journal} {Phys. Rev. B}\ }\textbf {\bibinfo {volume} {96}},\ \bibinfo {pages} {134313} (\bibinfo {year} {2017})}\BibitemShut {NoStop}%
\bibitem [{\citenamefont {\ifmmode \check{Z}\else \v{Z}\fi{}unkovi\ifmmode~\check{c}\else \v{c}\fi{}}\ \emph {et~al.}(2018)\citenamefont {\ifmmode \check{Z}\else \v{Z}\fi{}unkovi\ifmmode~\check{c}\else \v{c}\fi{}}, \citenamefont {Heyl}, \citenamefont {Knap},\ and\ \citenamefont {Silva}}]{Zunkovic18}%
  \BibitemOpen
  \bibfield  {author} {\bibinfo {author} {\bibfnamefont {B.}~\bibnamefont {\ifmmode \check{Z}\else \v{Z}\fi{}unkovi\ifmmode~\check{c}\else \v{c}\fi{}}}, \bibinfo {author} {\bibfnamefont {M.}~\bibnamefont {Heyl}}, \bibinfo {author} {\bibfnamefont {M.}~\bibnamefont {Knap}}, \ and\ \bibinfo {author} {\bibfnamefont {A.}~\bibnamefont {Silva}},\ }\href {\doibase 10.1103/PhysRevLett.120.130601} {\bibfield  {journal} {\bibinfo  {journal} {Phys. Rev. Lett.}\ }\textbf {\bibinfo {volume} {120}},\ \bibinfo {pages} {130601} (\bibinfo {year} {2018})}\BibitemShut {NoStop}%
\bibitem [{\citenamefont {Halimeh}\ \emph {et~al.}(2017)\citenamefont {Halimeh}, \citenamefont {Zauner-Stauber}, \citenamefont {McCulloch}, \citenamefont {de~Vega}, \citenamefont {Schollw\"{o}ck},\ and\ \citenamefont {Kastner}}]{Halimeh2017}%
  \BibitemOpen
  \bibfield  {author} {\bibinfo {author} {\bibfnamefont {J.~C.}\ \bibnamefont {Halimeh}}, \bibinfo {author} {\bibfnamefont {V.}~\bibnamefont {Zauner-Stauber}}, \bibinfo {author} {\bibfnamefont {I.~P.}\ \bibnamefont {McCulloch}}, \bibinfo {author} {\bibfnamefont {I.}~\bibnamefont {de~Vega}}, \bibinfo {author} {\bibfnamefont {U.}~\bibnamefont {Schollw\"{o}ck}}, \ and\ \bibinfo {author} {\bibfnamefont {M.}~\bibnamefont {Kastner}},\ }\href {\doibase 10.1103/physrevb.95.024302} {\bibfield  {journal} {\bibinfo  {journal} {Physical Review B}\ }\textbf {\bibinfo {volume} {95}} (\bibinfo {year} {2017}),\ 10.1103/physrevb.95.024302}\BibitemShut {NoStop}%
\bibitem [{\citenamefont {Lang}\ \emph {et~al.}(2018{\natexlab{a}})\citenamefont {Lang}, \citenamefont {Frank},\ and\ \citenamefont {Halimeh}}]{lang2018dynamical}%
  \BibitemOpen
  \bibfield  {author} {\bibinfo {author} {\bibfnamefont {J.}~\bibnamefont {Lang}}, \bibinfo {author} {\bibfnamefont {B.}~\bibnamefont {Frank}}, \ and\ \bibinfo {author} {\bibfnamefont {J.~C.}\ \bibnamefont {Halimeh}},\ }\href@noop {} {\bibfield  {journal} {\bibinfo  {journal} {Phys. Rev. Lett.}\ }\textbf {\bibinfo {volume} {121}},\ \bibinfo {pages} {130603} (\bibinfo {year} {2018}{\natexlab{a}})}\BibitemShut {NoStop}%
\bibitem [{\citenamefont {Lang}\ \emph {et~al.}(2018{\natexlab{b}})\citenamefont {Lang}, \citenamefont {Frank},\ and\ \citenamefont {Halimeh}}]{lang2018concurrence}%
  \BibitemOpen
  \bibfield  {author} {\bibinfo {author} {\bibfnamefont {J.}~\bibnamefont {Lang}}, \bibinfo {author} {\bibfnamefont {B.}~\bibnamefont {Frank}}, \ and\ \bibinfo {author} {\bibfnamefont {J.~C.}\ \bibnamefont {Halimeh}},\ }\href@noop {} {\bibfield  {journal} {\bibinfo  {journal} {Physical Review B}\ }\textbf {\bibinfo {volume} {97}},\ \bibinfo {pages} {174401} (\bibinfo {year} {2018}{\natexlab{b}})}\BibitemShut {NoStop}%
\bibitem [{\citenamefont {Halimeh}\ and\ \citenamefont {Zauner-Stauber}(2016)}]{halimeh2016dynamical}%
  \BibitemOpen
  \bibfield  {author} {\bibinfo {author} {\bibfnamefont {J.~C.}\ \bibnamefont {Halimeh}}\ and\ \bibinfo {author} {\bibfnamefont {V.}~\bibnamefont {Zauner-Stauber}},\ }\href@noop {} {\bibfield  {journal} {\bibinfo  {journal} {arXiv preprint arXiv:1610.02019}\ } (\bibinfo {year} {2016})}\BibitemShut {NoStop}%
\bibitem [{\citenamefont {Homrighausen}\ \emph {et~al.}(2017)\citenamefont {Homrighausen}, \citenamefont {Abeling}, \citenamefont {Zauner-Stauber},\ and\ \citenamefont {Halimeh}}]{homrighausen2017anomalous}%
  \BibitemOpen
  \bibfield  {author} {\bibinfo {author} {\bibfnamefont {I.}~\bibnamefont {Homrighausen}}, \bibinfo {author} {\bibfnamefont {N.~O.}\ \bibnamefont {Abeling}}, \bibinfo {author} {\bibfnamefont {V.}~\bibnamefont {Zauner-Stauber}}, \ and\ \bibinfo {author} {\bibfnamefont {J.~C.}\ \bibnamefont {Halimeh}},\ }\href@noop {} {\bibfield  {journal} {\bibinfo  {journal} {Physical Review B}\ }\textbf {\bibinfo {volume} {96}},\ \bibinfo {pages} {104436} (\bibinfo {year} {2017})}\BibitemShut {NoStop}%
\bibitem [{\citenamefont {Sciolla}\ and\ \citenamefont {Biroli}(2011)}]{Sciolla2011}%
  \BibitemOpen
  \bibfield  {author} {\bibinfo {author} {\bibfnamefont {B.}~\bibnamefont {Sciolla}}\ and\ \bibinfo {author} {\bibfnamefont {G.}~\bibnamefont {Biroli}},\ }\href {\doibase 10.1088/1742-5468/2011/11/p11003} {\bibfield  {journal} {\bibinfo  {journal} {Journal of Statistical Mechanics: Theory and Experiment}\ }\textbf {\bibinfo {volume} {2011}},\ \bibinfo {pages} {P11003} (\bibinfo {year} {2011})}\BibitemShut {NoStop}%
\bibitem [{\citenamefont {Jurcevic}\ \emph {et~al.}(2017)\citenamefont {Jurcevic}, \citenamefont {Shen}, \citenamefont {Hauke}, \citenamefont {Maier}, \citenamefont {Brydges}, \citenamefont {Hempel}, \citenamefont {Lanyon}, \citenamefont {Heyl}, \citenamefont {Blatt},\ and\ \citenamefont {Roos}}]{Jurcevic17}%
  \BibitemOpen
  \bibfield  {author} {\bibinfo {author} {\bibfnamefont {P.}~\bibnamefont {Jurcevic}}, \bibinfo {author} {\bibfnamefont {H.}~\bibnamefont {Shen}}, \bibinfo {author} {\bibfnamefont {P.}~\bibnamefont {Hauke}}, \bibinfo {author} {\bibfnamefont {C.}~\bibnamefont {Maier}}, \bibinfo {author} {\bibfnamefont {T.}~\bibnamefont {Brydges}}, \bibinfo {author} {\bibfnamefont {C.}~\bibnamefont {Hempel}}, \bibinfo {author} {\bibfnamefont {B.~P.}\ \bibnamefont {Lanyon}}, \bibinfo {author} {\bibfnamefont {M.}~\bibnamefont {Heyl}}, \bibinfo {author} {\bibfnamefont {R.}~\bibnamefont {Blatt}}, \ and\ \bibinfo {author} {\bibfnamefont {C.~F.}\ \bibnamefont {Roos}},\ }\href {\doibase 10.1103/PhysRevLett.119.080501} {\bibfield  {journal} {\bibinfo  {journal} {Phys. Rev. Lett.}\ }\textbf {\bibinfo {volume} {119}},\ \bibinfo {pages} {080501} (\bibinfo {year} {2017})}\BibitemShut {NoStop}%
\bibitem [{\citenamefont {Lerose}\ \emph {et~al.}(2018)\citenamefont {Lerose}, \citenamefont {Marino}, \citenamefont {\ifmmode \check{Z}\else \v{Z}\fi{}unkovi\ifmmode~\check{c}\else \v{c}\fi{}}, \citenamefont {Gambassi},\ and\ \citenamefont {Silva}}]{Lerose2018}%
  \BibitemOpen
  \bibfield  {author} {\bibinfo {author} {\bibfnamefont {A.}~\bibnamefont {Lerose}}, \bibinfo {author} {\bibfnamefont {J.}~\bibnamefont {Marino}}, \bibinfo {author} {\bibfnamefont {B.}~\bibnamefont {\ifmmode \check{Z}\else \v{Z}\fi{}unkovi\ifmmode~\check{c}\else \v{c}\fi{}}}, \bibinfo {author} {\bibfnamefont {A.}~\bibnamefont {Gambassi}}, \ and\ \bibinfo {author} {\bibfnamefont {A.}~\bibnamefont {Silva}},\ }\href {\doibase 10.1103/PhysRevLett.120.130603} {\bibfield  {journal} {\bibinfo  {journal} {Phys. Rev. Lett.}\ }\textbf {\bibinfo {volume} {120}},\ \bibinfo {pages} {130603} (\bibinfo {year} {2018})}\BibitemShut {NoStop}%
\bibitem [{\citenamefont {Lerose}\ \emph {et~al.}(2019)\citenamefont {Lerose}, \citenamefont {{\v{Z}}unkovi{\v{c}}}, \citenamefont {Marino}, \citenamefont {Gambassi},\ and\ \citenamefont {Silva}}]{Lerose2019}%
  \BibitemOpen
  \bibfield  {author} {\bibinfo {author} {\bibfnamefont {A.}~\bibnamefont {Lerose}}, \bibinfo {author} {\bibfnamefont {B.}~\bibnamefont {{\v{Z}}unkovi{\v{c}}}}, \bibinfo {author} {\bibfnamefont {J.}~\bibnamefont {Marino}}, \bibinfo {author} {\bibfnamefont {A.}~\bibnamefont {Gambassi}}, \ and\ \bibinfo {author} {\bibfnamefont {A.}~\bibnamefont {Silva}},\ }\href {\doibase 10.1103/physrevb.99.045128} {\bibfield  {journal} {\bibinfo  {journal} {Physical Review B}\ }\textbf {\bibinfo {volume} {99}} (\bibinfo {year} {2019}),\ 10.1103/physrevb.99.045128}\BibitemShut {NoStop}%
\bibitem [{\citenamefont {Piccitto}\ \emph {et~al.}(2019)\citenamefont {Piccitto}, \citenamefont {\ifmmode \check{Z}\else \v{Z}\fi{}unkovi\ifmmode~\check{c}\else \v{c}\fi{}},\ and\ \citenamefont {Silva}}]{Piccitto19}%
  \BibitemOpen
  \bibfield  {author} {\bibinfo {author} {\bibfnamefont {G.}~\bibnamefont {Piccitto}}, \bibinfo {author} {\bibfnamefont {B.}~\bibnamefont {\ifmmode \check{Z}\else \v{Z}\fi{}unkovi\ifmmode~\check{c}\else \v{c}\fi{}}}, \ and\ \bibinfo {author} {\bibfnamefont {A.}~\bibnamefont {Silva}},\ }\href {\doibase 10.1103/PhysRevB.100.180402} {\bibfield  {journal} {\bibinfo  {journal} {Phys. Rev. B}\ }\textbf {\bibinfo {volume} {100}},\ \bibinfo {pages} {180402} (\bibinfo {year} {2019})}\BibitemShut {NoStop}%
\bibitem [{\citenamefont {Piccitto}\ and\ \citenamefont {Silva}(2019)}]{Piccitto19b}%
  \BibitemOpen
  \bibfield  {author} {\bibinfo {author} {\bibfnamefont {G.}~\bibnamefont {Piccitto}}\ and\ \bibinfo {author} {\bibfnamefont {A.}~\bibnamefont {Silva}},\ }\href {\doibase 10.1088/1742-5468/ab3a27} {\bibfield  {journal} {\bibinfo  {journal} {Journal of Statistical Mechanics: Theory and Experiment}\ }\textbf {\bibinfo {volume} {2019}},\ \bibinfo {pages} {094017} (\bibinfo {year} {2019})}\BibitemShut {NoStop}%
\bibitem [{\citenamefont {Canovi}\ \emph {et~al.}(2014)\citenamefont {Canovi}, \citenamefont {Werner},\ and\ \citenamefont {Eckstein}}]{Canovi14}%
  \BibitemOpen
  \bibfield  {author} {\bibinfo {author} {\bibfnamefont {E.}~\bibnamefont {Canovi}}, \bibinfo {author} {\bibfnamefont {P.}~\bibnamefont {Werner}}, \ and\ \bibinfo {author} {\bibfnamefont {M.}~\bibnamefont {Eckstein}},\ }\href {\doibase 10.1103/PhysRevLett.113.265702} {\bibfield  {journal} {\bibinfo  {journal} {Phys. Rev. Lett.}\ }\textbf {\bibinfo {volume} {113}},\ \bibinfo {pages} {265702} (\bibinfo {year} {2014})}\BibitemShut {NoStop}%
\bibitem [{\citenamefont {Wang}\ and\ \citenamefont {Fazio}(2021)}]{wang2021dissipative}%
  \BibitemOpen
  \bibfield  {author} {\bibinfo {author} {\bibfnamefont {P.}~\bibnamefont {Wang}}\ and\ \bibinfo {author} {\bibfnamefont {R.}~\bibnamefont {Fazio}},\ }\href {\doibase 10.1103/physreva.103.013306} {\bibfield  {journal} {\bibinfo  {journal} {Physical Review A}\ }\textbf {\bibinfo {volume} {103}} (\bibinfo {year} {2021}),\ 10.1103/physreva.103.013306}\BibitemShut {NoStop}%
\bibitem [{\citenamefont {R\"{u}ckriegel}\ \emph {et~al.}(2012)\citenamefont {R\"{u}ckriegel}, \citenamefont {Kreisel},\ and\ \citenamefont {Kopietz}}]{ruckriegel2012time}%
  \BibitemOpen
  \bibfield  {author} {\bibinfo {author} {\bibfnamefont {A.}~\bibnamefont {R\"{u}ckriegel}}, \bibinfo {author} {\bibfnamefont {A.}~\bibnamefont {Kreisel}}, \ and\ \bibinfo {author} {\bibfnamefont {P.}~\bibnamefont {Kopietz}},\ }\href {\doibase 10.1103/physrevb.85.054422} {\bibfield  {journal} {\bibinfo  {journal} {Physical Review B}\ }\textbf {\bibinfo {volume} {85}} (\bibinfo {year} {2012}),\ 10.1103/physrevb.85.054422}\BibitemShut {NoStop}%
\bibitem [{\citenamefont {Derrida}(1980)}]{derrida1980random}%
  \BibitemOpen
  \bibfield  {author} {\bibinfo {author} {\bibfnamefont {B.}~\bibnamefont {Derrida}},\ }\href {\doibase 10.1103/physrevlett.45.79} {\bibfield  {journal} {\bibinfo  {journal} {Phys. Rev. Lett.}\ }\textbf {\bibinfo {volume} {45}},\ \bibinfo {pages} {79} (\bibinfo {year} {1980})}\BibitemShut {NoStop}%
\bibitem [{\citenamefont {Derrida}(1981)}]{derrida1981random}%
  \BibitemOpen
  \bibfield  {author} {\bibinfo {author} {\bibfnamefont {B.}~\bibnamefont {Derrida}},\ }\href {\doibase 10.1103/physrevb.24.2613} {\bibfield  {journal} {\bibinfo  {journal} {Physical Review B}\ }\textbf {\bibinfo {volume} {24}},\ \bibinfo {pages} {2613} (\bibinfo {year} {1981})}\BibitemShut {NoStop}%
\bibitem [{\citenamefont {J\"{o}rg}\ \emph {et~al.}(2010)\citenamefont {J\"{o}rg}, \citenamefont {Krzakala}, \citenamefont {Kurchan}, \citenamefont {Maggs},\ and\ \citenamefont {Pujos}}]{jorg2010energy}%
  \BibitemOpen
  \bibfield  {author} {\bibinfo {author} {\bibfnamefont {T.}~\bibnamefont {J\"{o}rg}}, \bibinfo {author} {\bibfnamefont {F.}~\bibnamefont {Krzakala}}, \bibinfo {author} {\bibfnamefont {J.}~\bibnamefont {Kurchan}}, \bibinfo {author} {\bibfnamefont {A.~C.}\ \bibnamefont {Maggs}}, \ and\ \bibinfo {author} {\bibfnamefont {J.}~\bibnamefont {Pujos}},\ }\href {\doibase 10.1209/0295-5075/89/40004} {\bibfield  {journal} {\bibinfo  {journal} {{EPL} (Europhysics Letters)}\ }\textbf {\bibinfo {volume} {89}},\ \bibinfo {pages} {40004} (\bibinfo {year} {2010})}\BibitemShut {NoStop}%
\bibitem [{\citenamefont {Bapst}\ and\ \citenamefont {Semerjian}(2012)}]{bapst2012quantum}%
  \BibitemOpen
  \bibfield  {author} {\bibinfo {author} {\bibfnamefont {V.}~\bibnamefont {Bapst}}\ and\ \bibinfo {author} {\bibfnamefont {G.}~\bibnamefont {Semerjian}},\ }\href {\doibase 10.1088/1742-5468/2012/06/p06007} {\bibfield  {journal} {\bibinfo  {journal} {Journal of Statistical Mechanics: Theory and Experiment}\ }\textbf {\bibinfo {volume} {2012}},\ \bibinfo {pages} {P06007} (\bibinfo {year} {2012})}\BibitemShut {NoStop}%
\bibitem [{\citenamefont {Das}\ \emph {et~al.}(2006)\citenamefont {Das}, \citenamefont {Sengupta}, \citenamefont {Sen},\ and\ \citenamefont {Chakrabarti}}]{das2006infinite}%
  \BibitemOpen
  \bibfield  {author} {\bibinfo {author} {\bibfnamefont {A.}~\bibnamefont {Das}}, \bibinfo {author} {\bibfnamefont {K.}~\bibnamefont {Sengupta}}, \bibinfo {author} {\bibfnamefont {D.}~\bibnamefont {Sen}}, \ and\ \bibinfo {author} {\bibfnamefont {B.~K.}\ \bibnamefont {Chakrabarti}},\ }\href {\doibase 10.1103/physrevb.74.144423} {\bibfield  {journal} {\bibinfo  {journal} {Physical Review B}\ }\textbf {\bibinfo {volume} {74}} (\bibinfo {year} {2006}),\ 10.1103/physrevb.74.144423}\BibitemShut {NoStop}%
\bibitem [{\citenamefont {Filippone}\ \emph {et~al.}(2011)\citenamefont {Filippone}, \citenamefont {Dusuel},\ and\ \citenamefont {Vidal}}]{filippone2011quantum}%
  \BibitemOpen
  \bibfield  {author} {\bibinfo {author} {\bibfnamefont {M.}~\bibnamefont {Filippone}}, \bibinfo {author} {\bibfnamefont {S.}~\bibnamefont {Dusuel}}, \ and\ \bibinfo {author} {\bibfnamefont {J.}~\bibnamefont {Vidal}},\ }\href {\doibase 10.1103/PhysRevA.83.022327} {\bibfield  {journal} {\bibinfo  {journal} {Phys. Rev. A}\ }\textbf {\bibinfo {volume} {83}},\ \bibinfo {pages} {022327} (\bibinfo {year} {2011})}\BibitemShut {NoStop}%
\bibitem [{\citenamefont {Sachdev}(2011)}]{sachdev2011quantum}%
  \BibitemOpen
  \bibfield  {author} {\bibinfo {author} {\bibfnamefont {S.}~\bibnamefont {Sachdev}},\ }\href {https://books.google.it/books?id=F3IkpxwpqSgC} {\emph {\bibinfo {title} {Quantum Phase Transitions}}}\ (\bibinfo  {publisher} {Cambridge University Press},\ \bibinfo {year} {2011})\BibitemShut {NoStop}%
\bibitem [{\citenamefont {Mu\~noz Arias}\ \emph {et~al.}(2020)\citenamefont {Mu\~noz Arias}, \citenamefont {Deutsch}, \citenamefont {Jessen},\ and\ \citenamefont {Poggi}}]{munoz2020simulation}%
  \BibitemOpen
  \bibfield  {author} {\bibinfo {author} {\bibfnamefont {M.~H.}\ \bibnamefont {Mu\~noz Arias}}, \bibinfo {author} {\bibfnamefont {I.~H.}\ \bibnamefont {Deutsch}}, \bibinfo {author} {\bibfnamefont {P.~S.}\ \bibnamefont {Jessen}}, \ and\ \bibinfo {author} {\bibfnamefont {P.~M.}\ \bibnamefont {Poggi}},\ }\href {\doibase 10.1103/PhysRevA.102.022610} {\bibfield  {journal} {\bibinfo  {journal} {Phys. Rev. A}\ }\textbf {\bibinfo {volume} {102}},\ \bibinfo {pages} {022610} (\bibinfo {year} {2020})}\BibitemShut {NoStop}%
\bibitem [{Note1()}]{Note1}%
  \BibitemOpen
  \bibinfo {note} {Our discussion can also be generalized to non-vanishing pre-quench values of the transverse field, $g_0>0$, following the approach of Ref. \cite {Sciolla2011}.}\BibitemShut {Stop}%
\bibitem [{\citenamefont {Keeling}\ \emph {et~al.}(2010)\citenamefont {Keeling}, \citenamefont {Bhaseen},\ and\ \citenamefont {Simons}}]{keeling2010collective}%
  \BibitemOpen
  \bibfield  {author} {\bibinfo {author} {\bibfnamefont {J.}~\bibnamefont {Keeling}}, \bibinfo {author} {\bibfnamefont {M.}~\bibnamefont {Bhaseen}}, \ and\ \bibinfo {author} {\bibfnamefont {B.}~\bibnamefont {Simons}},\ }\href@noop {} {\bibfield  {journal} {\bibinfo  {journal} {Phys. Rev. Lett.}\ }\textbf {\bibinfo {volume} {105}},\ \bibinfo {pages} {043001} (\bibinfo {year} {2010})}\BibitemShut {NoStop}%
\bibitem [{\citenamefont {Gambassi}\ and\ \citenamefont {Calabrese}(2011)}]{Gambassi2011}%
  \BibitemOpen
  \bibfield  {author} {\bibinfo {author} {\bibfnamefont {A.}~\bibnamefont {Gambassi}}\ and\ \bibinfo {author} {\bibfnamefont {P.}~\bibnamefont {Calabrese}},\ }\href {\doibase 10.1209/0295-5075/95/66007} {\bibfield  {journal} {\bibinfo  {journal} {{EPL} (Europhysics Letters)}\ }\textbf {\bibinfo {volume} {95}},\ \bibinfo {pages} {66007} (\bibinfo {year} {2011})}\BibitemShut {NoStop}%
\bibitem [{\citenamefont {Dutta}\ and\ \citenamefont {Bhattacharjee}(2001)}]{dutta2001phase}%
  \BibitemOpen
  \bibfield  {author} {\bibinfo {author} {\bibfnamefont {A.}~\bibnamefont {Dutta}}\ and\ \bibinfo {author} {\bibfnamefont {J.}~\bibnamefont {Bhattacharjee}},\ }\href@noop {} {\bibfield  {journal} {\bibinfo  {journal} {Physical Review B}\ }\textbf {\bibinfo {volume} {64}},\ \bibinfo {pages} {184106} (\bibinfo {year} {2001})}\BibitemShut {NoStop}%
\bibitem [{\citenamefont {Mori}\ \emph {et~al.}(2018)\citenamefont {Mori}, \citenamefont {Ikeda}, \citenamefont {Kaminishi},\ and\ \citenamefont {Ueda}}]{Mori2018preth}%
  \BibitemOpen
  \bibfield  {author} {\bibinfo {author} {\bibfnamefont {T.}~\bibnamefont {Mori}}, \bibinfo {author} {\bibfnamefont {T.~N.}\ \bibnamefont {Ikeda}}, \bibinfo {author} {\bibfnamefont {E.}~\bibnamefont {Kaminishi}}, \ and\ \bibinfo {author} {\bibfnamefont {M.}~\bibnamefont {Ueda}},\ }\href {\doibase 10.1088/1361-6455/aabcdf} {\bibfield  {journal} {\bibinfo  {journal} {Journal of Physics B: Atomic, Molecular and Optical Physics}\ }\textbf {\bibinfo {volume} {51}},\ \bibinfo {pages} {112001} (\bibinfo {year} {2018})}\BibitemShut {NoStop}%
\bibitem [{\citenamefont {Gring}\ \emph {et~al.}(2012)\citenamefont {Gring}, \citenamefont {Kuhnert}, \citenamefont {Langen}, \citenamefont {Kitagawa}, \citenamefont {Rauer}, \citenamefont {Schreitl}, \citenamefont {Mazets}, \citenamefont {Smith}, \citenamefont {Demler},\ and\ \citenamefont {Schmiedmayer}}]{Gring2012prethermalization}%
  \BibitemOpen
  \bibfield  {author} {\bibinfo {author} {\bibfnamefont {M.}~\bibnamefont {Gring}}, \bibinfo {author} {\bibfnamefont {M.}~\bibnamefont {Kuhnert}}, \bibinfo {author} {\bibfnamefont {T.}~\bibnamefont {Langen}}, \bibinfo {author} {\bibfnamefont {T.}~\bibnamefont {Kitagawa}}, \bibinfo {author} {\bibfnamefont {B.}~\bibnamefont {Rauer}}, \bibinfo {author} {\bibfnamefont {M.}~\bibnamefont {Schreitl}}, \bibinfo {author} {\bibfnamefont {I.}~\bibnamefont {Mazets}}, \bibinfo {author} {\bibfnamefont {D.~A.}\ \bibnamefont {Smith}}, \bibinfo {author} {\bibfnamefont {E.}~\bibnamefont {Demler}}, \ and\ \bibinfo {author} {\bibfnamefont {J.}~\bibnamefont {Schmiedmayer}},\ }\href {\doibase 10.1126/science.1224953} {\bibfield  {journal} {\bibinfo  {journal} {Science}\ }\textbf {\bibinfo {volume} {337}},\ \bibinfo {pages} {1318} (\bibinfo {year} {2012})}\BibitemShut {NoStop}%
\bibitem [{\citenamefont {Langen}\ \emph {et~al.}(2016)\citenamefont {Langen}, \citenamefont {Gasenzer},\ and\ \citenamefont {Schmiedmayer}}]{Langen2016prethermalization}%
  \BibitemOpen
  \bibfield  {author} {\bibinfo {author} {\bibfnamefont {T.}~\bibnamefont {Langen}}, \bibinfo {author} {\bibfnamefont {T.}~\bibnamefont {Gasenzer}}, \ and\ \bibinfo {author} {\bibfnamefont {J.}~\bibnamefont {Schmiedmayer}},\ }\href {\doibase 10.1088/1742-5468/2016/06/064009} {\bibfield  {journal} {\bibinfo  {journal} {Journal of Statistical Mechanics: Theory and Experiment}\ }\textbf {\bibinfo {volume} {2016}},\ \bibinfo {pages} {064009} (\bibinfo {year} {2016})}\BibitemShut {NoStop}%
\bibitem [{\citenamefont {Auerbach}(2012)}]{auerbach2012interacting}%
  \BibitemOpen
  \bibfield  {author} {\bibinfo {author} {\bibfnamefont {A.}~\bibnamefont {Auerbach}},\ }\href@noop {} {\emph {\bibinfo {title} {Interacting electrons and quantum magnetism}}}\ (\bibinfo  {publisher} {Springer Science \& Business Media},\ \bibinfo {year} {2012})\BibitemShut {NoStop}%
\bibitem [{Note2()}]{Note2}%
  \BibitemOpen
  \bibinfo {note} {This is true as long as the $g$ is below the spinodal point $g_{sp}$ from Eq.~\protect \eqref {eq:spinodal_point}, a condition always satisfied in our study.}\BibitemShut {Stop}%
\bibitem [{Note3()}]{Note3}%
  \BibitemOpen
  \bibinfo {note} {We observe that the right-hand side of Eq.~\protect \eqref {eq:eom_swdensity} is invariant under reflection $\phi \to \phi + \pi $ with respect to the $z$-axis, so that spin-wave emission are symmetric in the two wells for $p=2$.}\BibitemShut {Stop}%
\bibitem [{\citenamefont {Vulpiani}\ \emph {et~al.}(2009)\citenamefont {Vulpiani}, \citenamefont {Cecconi},\ and\ \citenamefont {Cencini}}]{vulpiani2009chaos}%
  \BibitemOpen
  \bibfield  {author} {\bibinfo {author} {\bibfnamefont {A.}~\bibnamefont {Vulpiani}}, \bibinfo {author} {\bibfnamefont {F.}~\bibnamefont {Cecconi}}, \ and\ \bibinfo {author} {\bibfnamefont {M.}~\bibnamefont {Cencini}},\ }\href@noop {} {\emph {\bibinfo {title} {Chaos: from simple models to complex systems}}},\ Vol.~\bibinfo {volume} {17}\ (\bibinfo  {publisher} {World Scientific},\ \bibinfo {year} {2009})\BibitemShut {NoStop}%
\bibitem [{ani()}]{animations}%
  \BibitemOpen
  \href@noop {} {}\bibinfo {howpublished} {See ancillary files \emph{Animated plot p=3.mpeg} and \emph{Animated plot p=4.mpeg}.}\BibitemShut {Stop}%
\bibitem [{Note4()}]{Note4}%
  \BibitemOpen
  \bibinfo {note} {Here $\protect \mathbf {k}$ is a $d$-dimensional vector if the lattice has dimensionality $d>1$}\BibitemShut {NoStop}%
\bibitem [{Note5()}]{Note5}%
  \BibitemOpen
  \bibinfo {note} {Up to replacing all the terms in the form of $J\delta ^{\alpha \beta }(t)$ with the more generic expression $\delta ^{\alpha \beta }(t) \equiv \DOTSB \sum@ \slimits@ _{\protect \mathbf {k}\protect \ne \protect \mathbf {0}} \Delta ^{\alpha \beta }_k(t) \protect \tilde {J}_\protect \mathbf {k}/ (Ns)$.}\BibitemShut {Stop}%
\end{thebibliography}%
\end{document}